\documentclass{aastex631}

\usepackage{multirow}
\usepackage{amsmath}
\usepackage{hhline}
\usepackage{array, makecell}
\usepackage{booktabs, tabularx}
\usepackage{comment}
\usepackage{enumitem}
\usepackage{mathrsfs}

\usepackage{float}
\usepackage{xcolor}
\newcounter{tr}
\ifnum \value{tr}>5

\newcommand{\deletedD}[1]{{\color{red} Damien - Deleted: } \sout{#1}}

\newcommand{\authorcommentD}[1]{{\color{blue} Damien - Comment :} {\color{blue} #1}}

\else

\newcommand{\deletedD}[1]{}

\newcommand{\authorcommentD}[1]{}

\fi

\begin{document}


\title{Unraveling the Origins of GRB X-ray Plateaus through a Study of X-ray Flares}

\shorttitle{X-ray flare properties}


\shortauthors{Dereli-B\'egu\'e et al.}

\author[0000-0002-8852-7530]{H. Dereli-B\'egu\'e}
\affiliation{Department of Physics, Bar-Ilan University, Ramat-Gan 52900, Israel}

\author[0000-0001-8667-0889]{A. Pe{'}er}
\affiliation{Department of Physics, Bar-Ilan University, Ramat-Gan 52900, Israel}

\author[0000-0003-4477-1846]{D. B\'egu\'e}
\affiliation{Department of Physics, Bar-Ilan University, Ramat-Gan 52900, Israel}

\author[0000-0002-9769-8016]{F. Ryde}
\affiliation{Department of Physics, KTH Royal Institute of Technology and The Oskar Klein Centre, SE-106 91 Stockholm, Sweden}


\begin{abstract}

The X-ray light curves of gamma-ray bursts (GRBs) display complex features, including plateaus and flares, that challenge theoretical models.
Here, we study the properties of flares that are observed in the early afterglow phase (up to a few thousands of seconds). We split the sample 
into two groups: bursts with
and  without X-ray plateau.  
We find that the distributions of flare properties are similar in each group.
Specifically, the peak time ($t_{\rm pk}$) of the flares and the ratio of the flare width to the flare peak time ($w/t_{\rm pk}$) which is found to be $\approx 1$, regardless of the presence of a plateau.
We discuss these results in view of the different theoretical models aimed at explaining the origin of the plateau.
These results are difficult to explain by viewing angle effects or late time energy injection, but do not contradict
the idea that GRBs with X-ray plateau have low Lorentz factor, of the order of tens.  
For these GRBs, the dissipation processes that produce the flares naturally occur at smaller radii compared to GRBs with higher Lorentz factors, while the flares maintain a similar behaviour. Our results therefore provide an independent support for the idea that many GRBs have Lorentz factor of a few tens rather than a few hundreds.

\end{abstract}



\section{Introduction} \label{sec:intro}

Gamma-ray bursts (GRBs) are extremely energetic events and are also known to have highly relativistic
jets, with initial expansion Lorentz factors typically ranging from tens to thousands \citep{KP91, Fenimore93, Racusin11, DPR22}.
GRBs display two sequential phases: the prompt emission phase, followed by the afterglow phase. These two phases can be
explained within the framework of the classical GRB fireball model \citep{RM92, MRW98, Pir04, Mes06, KZ15} where the GRB prompt
$\gamma$-ray emission is caused by internal energy dissipation (e.g., shocks) within a collimated ultrarelativistic jet, and the
broadband afterglow emission is created by an external shock propagating into the circumburst medium, which can be either a
stellar wind or constant density interstellar medium (ISM) \citep{MR97, SPN98}. 

Following the launch of the Neil Gehrels \textit{Swift} Observatory (hereafter \textit{Swift}; \citet{Gehrels+04}), several
previously unknown features in the X-ray light curves have been observed. Notably, while the late-time X-ray emission (after
$\sim 10^{3.5}- 10^4$~s), aligns with the theoretical predictions of the classical fireball
model, other phenomena such as early steep decays, X-ray plateaus, and X-ray flares observed by \textit{Swift} do not fit as
straightforwardly within this model \citep{Nousek06, OBrien06, Zhang06, Evans2007, Evans09}.

The early steep decay, characterized by a temporal index between $3 \leq \alpha \leq 5$, has been associated with the end of
the prompt phase and is attributed to high-latitude emission arising from photons emitted at larger angles relative to the
jet axis (corresponding to the observer direction), leading to a rapid decline in observed intensity \citep{BCG05, TGC+05, WOO07, ROB+21}. 
The following X-ray plateaus, with a shallow temporal index between $0 \leq \alpha \leq 0.7$,
generally appear $100-10^3$ seconds after the GRB trigger and are usually followed by a late time "afterglow" decay with an index between
$1.2 \leq \alpha \leq 1.5$ \citep{Zhang06, DPR22}, as is expected theoretically \citep{MR93}. Therefore, the flux
during the plateau phase decreases slower than theoretically expected. This feature is observed in a substantial fraction, close to 60\% of all GRBs  \citep{Evans09, Srinivasaragavan20}.

Since its discovery in early 2005, several ideas have been proposed to explain the X-ray plateau phase. It was initially
proposed that the origin of the plateau phase is a continuous energy injection into the external shock from the central
compact object, which produces the energy powering the GRB. This central engine can either be a newly formed black-hole
\citep{Zhang06, GK06, Nousek06} or a millisecond magnetar \citep{Metzger11}. Alternatively, it was suggested that the
observed plateau is due to viewing angle effects, where structured jets are observed slightly off-axis
\citep{EG06, Toma06, Eichler14, BDD+20a, OAB+20, BGG20b}. Recently, we suggested a different explanation, according to which
the observed signal originates from an outflow that is observed on-axis, but reaches a maximum Lorentz factor of several
tens at most, expanding into a medium composed of a low density wind \citep{ShM12, DPR22}. It is therefore of interest
to find an observational measure that could discriminate between these models.

X-ray flares typically occur $100-10^5$ seconds after the prompt emission and are observed in about half of the GRB population,
mostly in long GRBs but rarely in short GRBs \citep{Burrows+05, Falcone+07, Chincarini+07,  Curran+08, Chincarini+10, MBB11, Bernardini+11}.
They usually appear as one or two flares, with cases of multiple flares being rare \citep[\textit{e.g.}][]{Perri+07, Abdo+11}. Flares
can be considerably energetic and are often characterized by large flux variations \citep{Falcone+06, Gibson+18}. Indeed, their fluence
can be up to $100\%$ of the prompt fluence, and the flare fluxes, measured with respect to the underlying continuum,
$\Delta F_{\rm flare}/F_{\rm cont.}$, can vary over short timescales $\Delta t/t_{pk} \lesssim 1$ where $\Delta t$ measures the duration of
the flare and $t_{pk}$ is the time of maximum flare flux with respect to the trigger time \citep{Chincarini+07}.  Since flares
share many properties with the prompt emission, it is widely accepted that they are powered by the late central engine
activities either by internal shocks \citep{IKZ05, FW05, Zhang06}, or some other dissipation process within the ultra-relativistic outflow
\citep{Giannios06, Lazzati+11}, but at later times and at lower energies.

Since flares are so abundant, they are observed both in GRBs with and without plateaus. Given that different theoretical
models about the origin of the plateau phase have different expectations about the observed properties of flares, comparing
flare properties in GRBs with and without plateaus can potentially be used as an independent way of discriminating between the
different models. The only study we aware of to day is that of \citet{Yi+22}, who compared the X-ray flare energy and the X-ray plateau
energy with the isotropic prompt emission energy and found that all are correlated. Here, we conduct a comprehensive study on
the properties of flares observed in \textit{Swift} GRBs with and without plateaus. We then discuss the implications on the
different theoretical models. 

This paper is organized as follows. In Section \ref{sec:select_sample}, we define the data collection and analyses method with
sample selection and model definition as well as fitting procedure and best model selection. In Section \ref{sec:results}, we
present the flare analysis results. In Section \ref{sec:discussion}, we then discuss how our definition of flares could impact the results. In Section
\ref{physical_interpretation}, we discuss the physical interpenetration of the results. Finally, in Section \ref{sec:conclusion}
we list our summary and conclusions. Throughout the paper, a flat $\Lambda$CDM cosmological model with cosmological parameters $\Omega_{\rm m} = 0.286$
and $H_0 = 70$ km.s$^{-1}$.Mpc$^{-1}$ are used \citep{Hinshaw+09}.


\section{Data Collection and Analysis Method} \label{sec:select_sample}

\subsection{Sample Selection}

We selected a statistically significant sample of 100 GRBs by anti-chronologically
searching through 8 years of data from the \textit{Swift} archive, starting December  2$^{\rm nd}$  2022.
As for our first paper on this topic \citep{DPR22}, the selection criteria are as follow. First, the redshift of the burst has
to be measured by spectroscopy. The second criteria limits the sample to those GRBs which triggered \textit{Swift}-BAT. Indeed,
GRBs which are not observed by BAT usually lack \textit{Swift}-XRT observations, preventing us from performing the fitting
procedure described in the following sections. In fact, even GRBs initially detected by \textit{Swift}-BAT can lack a sufficiently
large number of data points, and those GRBs were removed from the sample. The required number of data points is set to be $\geq$~5
at the beginning or end of each independent decaying feature identified in the light curve.

\subsection{Model Definition} \label{sec:model}

GRB afterglow light-curves are made of several components, namely (i) the early afterglow steep decay; (ii) the plateau
(when it exist); (iii) the late time afterglow ("the self-similar decay") slope; and (iv) the post-jet break decay.  On
top of these, there are (v) flares (when they exist). Combining all components lead to the so-called canonical X-ray light-curve
\citep{Nousek06, OBrien06, Zhang06}. Yet each component can be observed independently of the others, creating the rich afterglow
phenomena, and providing a unique challenge in fitting and interpreting the afterglow light-curves.

To investigate both flare and plateau properties, our analysis method is as follows. In our approach, we model the
light curve using continuous power-law segments, that can have either one, two, three or four segments. Flares
are modeled using the "Norris" function \citep{NBK05}. They are integrally part of the models and their properties are constrained
during the fit, alongside those of the underlying afterglow. For full details of the fitting procedure, see section
\ref{sec:Fitting_procedure} and Appendix~\ref{app:model_definition} below.   

The difficulty in handling multi-component models, considering for instance several flares, the underlying afterglow emission and
eventually a jet break, is due to the large number of parameters and the resulting high model flexibility, which allows the "naive"
model to fit any light-curve at the expense of a physical interpretation. A solution to this problem is to prescribe conditions that
are simultaneously sufficiently general for the model to accurately represent the data or clearly fail when attempting to do so,
while also being sufficiently physically motivated to obtain meaningful and interpretable results.

Here, we choose to fit each X-ray light curve, starting from the "steep decay" phase (when it exists). Following the steep decay,
we apply a physically motivated afterglow model, based on synchrotron emission from particles accelerated by the propagating forward
shock wave, assuming spherical symmetry. The shock-accelerated electrons assume a power-law distribution, with power law index $p$,
in between Lorentz factors $\gamma_{\min}$ and $\gamma_{\rm max}$. This leads to two relevant characteristic frequencies in the observed
spectra, $\nu_{\rm m}$ which is the typical emission frequency from electrons at $\gamma_{\min}$, and $\nu_{\rm c}$ which
is the cooling frequency. As was shown by several authors \citep[\textit{e.g.}][]{GS02, DPR22}, various combinations of light curves and
spectra can be obtained. The resulting light curve depends on (i) the evolutionary stage of the blast wave propagation - coasting vs.
self similar decaying phase; (ii) the ambient medium profile- stellar wind or constant density ISM; and (iii) the observed frequency,
which can be smaller, larger or in between $\nu_{\rm m}$ and $\nu_{\rm c}$. 

In addition to the above mentioned possibilities, when fitting the light curves, we add three additional categories: (i) A steep
decay is fitted as a power law in time, with a slope that is independent on the electrons power law index and steeper than 2; (ii)
inclusion of a jet break; and (iii) number of flares, which we consider to vary in between 0, 1 or 2. Combined together, we consider
a total of 36 different models used in fitting the data. Although the models we use assume spherical symmetry, we point out that
deviation from spherical symmetry predictions are expected at late times (after the jet break), while all the flares identified occur
at much earlier times, before the jet break. Therefore, the key properties of the flares  we are interested in (peak time and flare width)
are only weakly sensitive to this assumption. For a complete description of the models used, see Appendix~\ref{app:model_definition}.

In our analysis, flares take the shape of a Norris function \citep{NBK05, Chincarini+10}, 
\begin{equation}
\mathcal{N} (T) = A_{\rm f} ~\lambda \exp \left [ -\frac{\tau_1}{T - t_0} -\frac{T - t_0}{\tau_2}  \right ],
\end{equation}
if $T> t_0$ and $\mathcal{N} (T) = 0$ otherwise. Here, $A_{\rm f}$ is the flare amplitude, $t_0$ is the onset time of the flare,
$\tau_1$ and $\tau_2$ are two flare shape parameters related to the rise and decay phases altering the flare profile and in particular
its asymmetry, and $\lambda = \exp (2 \mu )$ where  $\mu = (\tau_1/\tau_2)^{1/2}$. The maximum of the flare is reached at time
$t_{\rm pk} = t_0 + (\tau_1 \tau_2)^{1/2}$ and its maximum equals to $\mathcal{N}(t_{\rm pk}) = A_{\rm f}$.  The factor $A_{\rm f}$
quantifies the relative enhancement of the flux due to the flare. Following \citet{Chincarini+10} and noting that we fit the logarithm
of the flux, one can define the flare flux variability,
\begin{equation}
\frac{\Delta F_{\rm flare}}{F_{\rm cont.}} = \frac{10^{\mathcal{N}(t_{\rm pk})} 10^{F_{\rm cont.}} - 10^{F_{\rm cont.}}}{10^{F_{\rm cont.}}} 
= 10^{\mathcal{N}(t_{\rm pk})} - 1.
\end{equation}
where $F_{\rm flare}$ and $F_{\rm cont.}$ are the flare and underlying afterglow fluxes.

An important property of X-ray flares is their temporal aspect ratio $ w/t_{\rm pk}$, where $w$ is the flare temporal width. Flare width
has been estimated in various ways and using different functions, including e.g. a symmetric Gaussian \citep{Chincarini+07}, a smoothly
broken power-law profile \citep{Yi+16}, and using the Norris function \citep{NBK05}. For the latter,
\citet{Chincarini+10, Bernardini+11} define the flare width $w$ as the time interval during which the contribution of the flare is larger
than $1/e$ of its maximum, leading to $w = \Delta t_{1/e} = \tau_2(1+4  \mu)^{1/2}$. We cannot use this definition directly, since the fits
are performed in log-space. Instead, we determine the times $\bar t_1$  and $\bar t_2 > \bar t_1$\footnote{We note these times $\bar t$ to
avoid confusion with the break time of the afterglow $T_1$, $T_2$ and $T_3$}, between which the Norris function has value at least half of
its maximum. The width is then define as
\begin{equation}
    w = 10^{\bar t_2} - 10^{\bar t_1},
\end{equation}
Furthermore, it is important to note that different definitions of flare widths can lead to variations in absolute width
measurements. For instance, the Full Width at Half Maximum (FWHM) may provide a narrower estimate compared
to widths calculated at lower thresholds, such as $5\%$ of the maximum flux. However, the overall
characterization of $w/t_{\rm pk}$ remains consistent across these definitions. Defining $w$ at $5\%$
of the maximum captures a broader interval around the peak than the FWHM,  while still providing
a similar interpretation in temporal analyses, particularly when comparing the relative timescales of flares.

Moreover, while the choice of model — such as the asymmetric Norris profile \citep{NBK05} — may produce different
width measurements compared to symmetric Gaussian profiles \citep{Chincarini+07}, these alternative definitions
do not substantially affect the theoretical interpretations of physical processes. Thus, although definitions
differ, the temporal ratio $w/t_{\rm pk}$ provides a robust measure across various models, allowing for a
consistent interpretation of flare variability and duration.

The isotropic-equivalent energy $E_{\rm iso, f}$ emitted in each flare can be calculated once the fit is performed. We calculate it using
\begin{equation}
    E_{\rm iso, f} = 4 \pi d_L^2  \cdot \mathrm{C_F} \cdot \int_{\bar t_1}^{\bar t_2} 10^{F_{cont.}} \left ( 10^{\mathcal{N}(T)} - 1 \right ) 10^T \ln(10)   \, dT. 
\end{equation}
Here, $d_L$ is the luminosity distance and $\mathrm{C_F}$ is the flux conversion factor,
which converts count rate to flux (obtained from the online \textit{Swift} repository). 
The integral represents the total count rate of the flare between $\bar t_1$ and $\bar t_2$. We point out that the total energy is not sensitive to the exact choice of the integration boundaries, as the flux rises and decays exponentially.

Finally, for comparison with previous studies \citep[\textit{e.g.}][]{NBK05, Chincarini+10}, we define the flare asymmetry, 
\begin{equation}
k = (1+4  \mu)^{-1/2}.
\label{eq:flare_asymetry}
\end{equation}

\subsection{Fitting procedure} \label{sec:Fitting_procedure}

The X-ray count rate light curves (hereinafter LCs) of each GRB have been downloaded from the online 
\textit{Swift} repository\footnote{\url{https://www.swift.ac.uk/xrt_curves/}} \citep{Evans2007, Evans09},
considering the full \textit{Swift}-XRT bandpass, \textit{i.e.} ($E_{\rm min}, E_{\rm max}$) = (0.3, 10) keV.
We then fit each individual LCs to all 36 considered models, all described in section \ref{sec:model} and
Appendix~\ref{app:model_definition}. Although the light curves could have been sorted into categories
in order to reduce the number of models per light-curve, we decided against this to avoid introducing human
bias into the analysis process and best model selection.   

To perform the fit, we assume a Gaussian likelihood and sample the posterior distributions with MultiNest \citep{FHB09},
a nested sampling algorithm designed for efficient Bayesian inference. We assume 8000 (1000 for the model without flares)
active points and a tolerance of $0.5$ to ensure efficient sampling and convergence. MultiNest offers a number of advantages,
including computational efficiency and the ability to robustly handle multi-modal posterior distributions, which is a
relevant possibility given the high dimensionality and complexity of the parameter space for the models we considered,
specifically for those with one or two flares. Multinest also computes the evidence, which is the key to perform model
comparison, see section \ref{sec:model_comp}.

The parameter ranges and their sampling distributions, either uniform or log-uniform, are summarized in Table
\ref{tab:param_priors}. For example, the electron power-law index $p$ is sampled linearly from a non-informative uniform
prior within the range of 2 to 3, similar to the parameters $A$, $b$, $q$, $A_{\rm f1}$, $A_{\rm f2}$ as seen in Table
\ref{tab:param_priors}. Parameters related to time \textit{i.e.}, $T_1$, $T_2$, $T_3$ as well as the rise and decay time
parameter for the Norris function for the flares follow log-uniform priors to account for their wide range and scale sensitivity.

In our analysis, for the flare amplitude, we set a minimum value of $0.2$ for $A_{f1}$, $A_{f2}$, corresponding to an increase
by a factor of 1.58 for the flare flux relative to the underlying continuum in the linear scale. This ensures that the flare flux
is always at least 1.58 times larger than the underlying continuum flux, allowing for clear identification of flares and preventing
the misidentification of small random flux variations in the light curve as flares.

\begin{table}[t]
\centering
\small
\begin{tabular}{c|c|c|c|c}
\hline
Parameter & Units  & Minimum & Maximum & Type of distribution \\
\hline
$T1$ & s & xdata(min) & 4 &  Logarithmic  \\
$T2$ & s & 0.5 &  3  & Logarithmic \\
$T3$ & s & 0.5 & 3 & Logarithmic \\
$A$ & - & -10 & 10  & Linear \\
$b$ & - & -5 & -2  & Linear \\
$p$ & - & 2 & 3 &  Linear  \\
$q$ & - & 0 & 1.5 &  Linear \\
\hline
$t_1$ & s & xdata(min)-0.1 & 3.5 & Logarithmic \\
$A_{f1}$ & - & 0.2 &  3 & Linear \\
$\tau_{1,1}$ & s & -2.1 &  -0.2  & Logarithmic  \\
$\tau_{1,2}$ & s & -2.1 &  -0.2 & Logarithmic  \\
$t_2$ & s & 0.2 & 2 &  Logarithmic   \\
$A_{f2}$ & - & 0.2 & 3  &  Linear \\
$\tau_{2,1}$& s & -2.1   & -0.2 & Logarithmic \\
$\tau_{2,2}$& s & -2.1 & -0.2  & Logarithmic \\
\hline
\end{tabular}
\caption{Parameters used in the fitting procedure. For each parameter we include its units, prior allowed ranges and the sampling distribution.
The parameters are divided into two parts. The top part includes the parameters of the underlying afterglow: $T_1$, $T_2$, and $T_3$
represent the break times; $A$ is the normalization; $b$ is the slope of the steep decay; $p$ is the electron power-law index; and $q$, together with $p$,
defines the slope of the decay phase after the jet break (see Appendix \ref{app:model_definition}). The bottom part includes the parameters of the flares when present: $t_1$ and $t_2$ are the start times of the flares; $A_{f1}$ and $A_{f2}$ are the flare amplitudes; and $\tau_{1,1}$, $\tau_{1,2}$,
$\tau_{2,1}$, and $\tau_{2,2}$ are the rise and decay time parameters that shape the flare profiles.}
\label{tab:param_priors} 
\end{table}

\subsection{Best model selection and verification} \label{sec:model_comp}

Our best model selection is based on the Bayesian model comparison methods and the following strategy. The first step
is to discard cases where there is a lack of data near the peak time of the flare. One of the challenges in analyzing GRB
X-ray data is the lack of observations, specifically on the time scale of a few thousands seconds, corresponding to the
orbit time of Swift (\textit{e.g.}, Earth occultation), which can significantly affect the accuracy of the model fitting.
To address this, we incorporate a criterion that considers the density of data points around the peak time of each flare.
Specifically, if there are fewer than two data points within the time interval from the start to the peak of the flare,
the model is flagged as unreliable and the model is discarded from further consideration. Indeed, this situation arises when
a single or two data points within the full LC can be better explained by a very narrow flare. In fact, this criterion
ensures that the model has sufficient data to accurately characterize the rise, the peak and the decay of the flare.

In the next step, we aim to balance the complexity of the model with its ability to fit the data well. This is achieved by
prioritizing models with fewer parameters, provided they adequately capture the observed phenomena. The potential best models
are identified for each case (C/F and E for the wind or H and G for the ISM, see Appendix \ref{app:model_definition} for details on the nomenclature)
by comparing nested models one by one based on the Bayes factors and excluding those with a Bayes factor larger than 5.
This process ensures that only models with balanced performance and fewer parameters are advanced to the next stage. 
While this initial selection narrows down the models effectively, Bayes factors are not always well-suited for non-nested model comparisons,
as prior choices can be haevily influence the Bayes factor.
To address this limitation, we adopt AIC as an additional criterion. AIC and AICc (corrected Akaike Information Criterion) provide a consistent
framework for comparing non-nested
models. Therefore, among all potential models, the best model is selected based on the lowest AICc value, as AICc accounts for
potential over-fitting and bias, particularly in models with small sample sizes. If AICc is unavailable, the model with the lowest AIC is chosen instead.
Finally, the residuals between the model and the data are computed for each selected model and visually inspected to ensure that it does not
exhibit outliers — namely data points with residuals significantly different from the others — or systematic deviations, which are consistent patterns of misfit.
Once the fit are finished and the analysis performed, we exclude 11 GRBs from the remainingof the analysis, as in these GRBs flares were misidentified,
resulting in failures of the different afterglow models to adequately fit the data. A detailed description of the reasoning for excluding these eleven GRBs are provided in Appendix~\ref{app:exluded_bursts}.


\section{Flare analysis results} \label{sec:results}

From the remaining 89 bursts in our sample, we find the following.
\begin{enumerate}[label=(\roman*)]
\item  61 bursts (69\% of  GRBs
in our sample) have flares. This indicates that flares are quite common in GRB afterglows. This value is slightly higher than previously reported, where about half of the GRBs were found to exhibit flares \citep{Zhang06, Chincarini+10}.
\item 57 GRBs in our sample (64\%) have an X-ray plateau. 
\item Among the 61 GRBs with flares, 42 (68\%) have a plateau, while 19 bursts do not.
\item Of the 57 GRBs that have a plateau, 42 (73\%) also exhibit flares. 
\end{enumerate}
These results imply that the overall occurrence of flares in GRBs without a plateau phase (19/32 = 59\%) does not statistically
differ from the occurrence of flares in GRBs with a plateau phase (42/57 = 73\%). This strongly suggests that the presence of
flares is independent of the existence of a plateau, indicating that these two phenomena, namely plateaus and flares are most
likely not related or dependent on each other. 

Here, we focus on analyzing the properties of flares, and we therefore focus on the sub-sample of 61 GRBs with flares. A complete analysis
of the afterglow properties of all GRBs in our sample will appear elsewhere (Dereli-B\'egu\'e et al., in prep.). The properties of all flares
that appear in the 61 bursts in our sample are listed in Table~\ref{tab:derived_params1} in Appendix~\ref{app:GRBs_with_flares_derived_params}.

\subsection{Flare energetic, time of occurrence and duration} \label{sec:flare}

\begin{figure}
\centering
\includegraphics[width=0.99\linewidth]{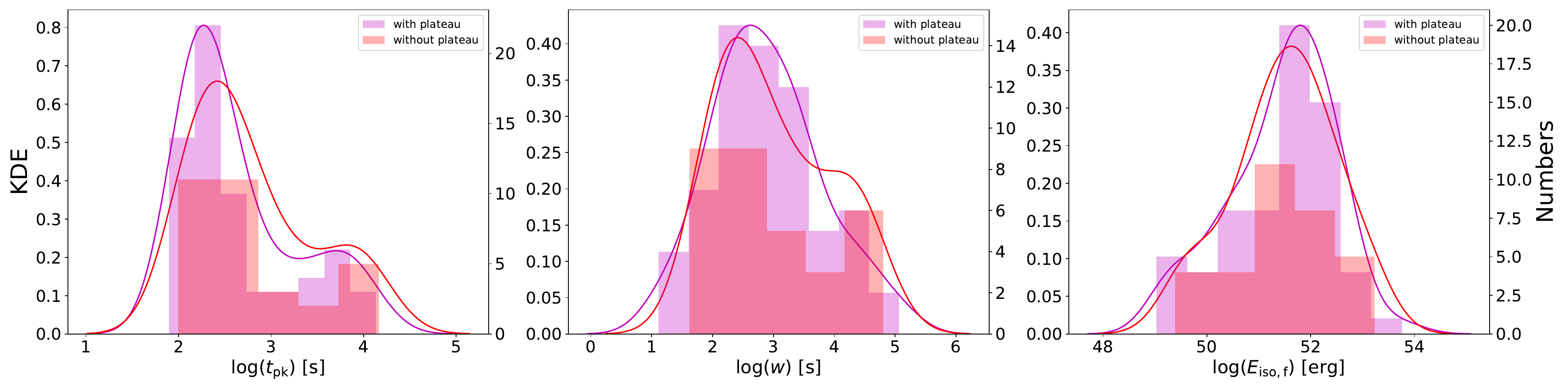}
\caption{Distributions of the flare peak time, $t_{\rm pk}$ (left),  the flare width $w$ (middle), and the flare isotropic energy,
$E_{\rm iso,f}$ (right). The 65 flares obtained from the 42 bursts with a plateau phase are in purple, while the 32 flares obtained
from the 19 GRBs without a plateau phase are in red. In each panel, the right-hand ordinate shows the
number of bursts in each bin while the left-hand ordinate shows the value of the kernel density estimation (KDE) drawn by the purple
and red solid lines. These distributions show that the properties of the flare are independent of the presence of a
plateau phase in the GRBs X-ray light curves.}
\label{fig:hist_peak-time_width_Eiso-flare}
\end{figure}

In order to study possible correlations between flare properties and existence of a plateau, we split the flare sample into two.
The first part contains 65 flares identified in the X-ray LCs of 42 bursts with a plateau phase, while the second part includes 32
flares identified in the LCs of 19 GRBs without a plateau phase. Figure \ref{fig:hist_peak-time_width_Eiso-flare} shows the distributions
of the flare peak times $t_{\rm pk}$, the flare width $w$, and the flare isotropic energy $E_{\rm iso,f}$. Flares that originate from
bursts with plateau are colored in purple, and those that are detected in GRBs without plateau are in red. Comparing the averages,
we find that $\langle \log_{10} t_{\rm pk} \rangle = 2.64 \pm 0.08$ for GRBs with a plateau, to be compared with
$\langle \log_{10} t_{\rm pk} \rangle = 2.79 \pm 0.11$ for burst without a plateau. A similar result holds for the width $w$ and
the flare isotropic energy $E_{\rm iso,f}$, with $ \langle \log_{10} w \rangle = 2.89 \pm 0.11$ and
$\langle \log_{10} E_{iso,f} \rangle = 51.43 \pm 0.12$ to be compared to $\langle \log_{10} w \rangle = 3.02 \pm 0.16$ and
$\langle \log_{10} E_{iso,f} \rangle = 51.49 \pm 0.17$ for bursts with and without plateau, respectively. We further performed
a Kolmogorov-Smirnov (KS) test for each parameter and find that (i) KS test statistic is $D = 0.23$ with a probability $p = 0.18$
for $t_{\rm pk}$; (ii) $D = 0.11$ and $p = 0.92$ for the width; and (iii) $D = 0.10$ and $p = 0.96$ for $E_{\rm iso,f}$. These results
show that the two sub-samples originate from the same population. We therefore conclude that the flares origin and properties are
independent on the afterglow properties, which are determined by the ambient medium and forward shock properties. In particular, this
means that flares are independent of the existence of a plateau phase. This conclusion is aligned with the conclusion of
\citet{Chincarini+10}, who similarly argued that flare properties are independent of the ambient medium (though plateaus were not
considered in that work).  

Furthermore, from Figure \ref{fig:hist_peak-time_width_Eiso-flare}, we find that both sub-samples show a heavy tail toward high
flare peak times and widths. This may indicate a possible existence of two populations of flares, namely narrow and wide flares.
We point out that the current statistics in this region of the parameter space is small. An analysis of the complete sample of
bursts observed with \textit{Swift} would allow to resolve this issue and firmly confirmed the existence of two kinds of flares.

\begin{figure}
\centering
\includegraphics[width=0.45\linewidth]{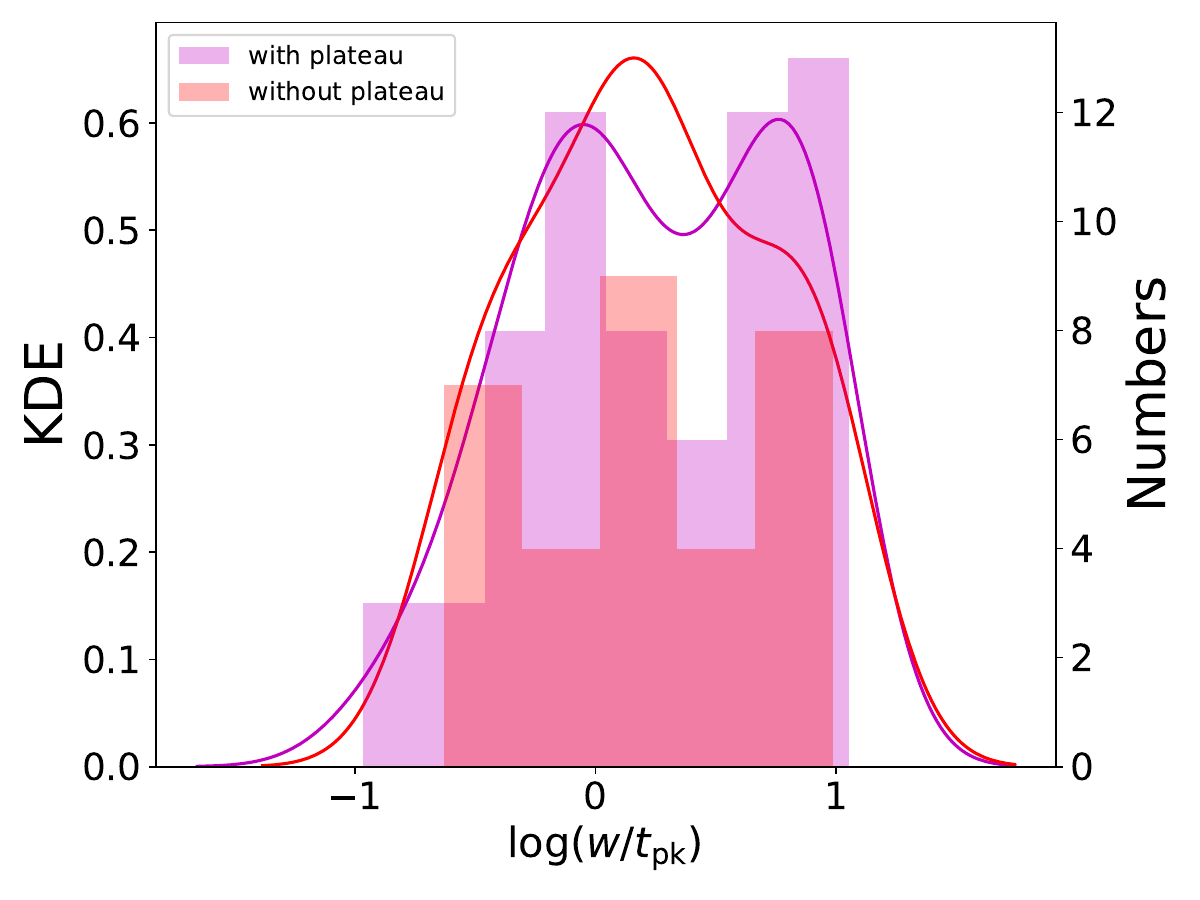}
\includegraphics[width=0.45\linewidth]{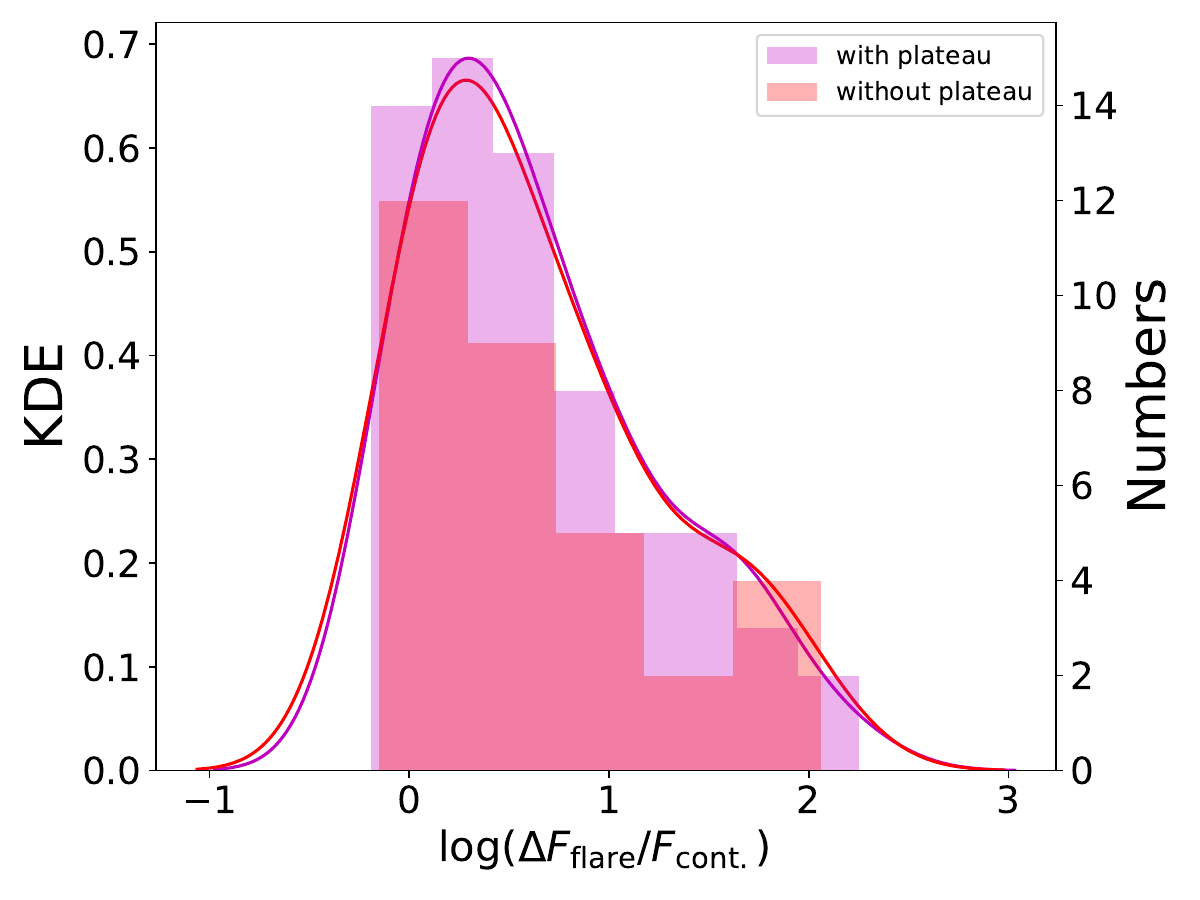}
\caption{Distributions of the ratio of the flare width to the flare peak time, $w/t_{\rm pk}$ (left), and the flare flux variability,
$\Delta F_{\rm flare}/F_{\rm cont.}$ (right). The data presentation, including the color coding, is the same as in Figure
\ref{fig:hist_peak-time_width_Eiso-flare}, where purple represents the 65 flares from 42 bursts with a plateau phase, and
red represents the 32 flares from 19 GRBs without plateau phases. The errors correspond to a significance of one sigma.
The results show that the flare time and flux variability properties are the same, regardless of whether there is a plateau
phase in the GRBs X-ray light curves. Additionally, bimodal distributions are observed in both sub-samples in the distributions
of the ratio of the flare width to the flare peak time. This bi-modality is more pronounced for the GRBs with a plateau phase.} 
\label{fig:hist_time_flux_varabilities}
\end{figure}

In Figure \ref{fig:hist_time_flux_varabilities}, we present the distribution of the ratio of the flare width to the
flare peak time $ w /t_{\rm pk}$ and the distribution of the flare flux variability $\Delta F_{\rm flare}/F_{\rm cont.}$.
Comparing the averages, we find that $\left < \log_{10} w/t_{\rm pk}\right > = 0.25 \pm 0.07$ for GRBs with a plateau, to
be compared with $\left < \log_{10} w/t_{\rm pk}\right > = 0.22 \pm 0.09$ for burst without a plateau. A similar result
holds for the flux variability, $\left < \log_{10} \Delta F_{\rm flare}/F_{\rm cont.}\right > = 0.63 \pm 0.07$ to be compared
with $\left < \log_{10} \Delta F_{\rm flare}/F_{\rm cont.}\right > = 0.63 \pm 0.11$ for bursts with and without a plateau,
respectively. A KS test for each parameter reveals $D = 0.12$ and $p = 0.88$ for $w/t_{\rm pk}$, and $D = 0.08$ and $p = 0.99$
for $\Delta F_{\rm flare}/F_{\rm cont.}$. These results again show that the two sub-samples originate from the same population.
An important result is that the ratio of the flare width to the flare peak time ($w/t_{\rm pk}$) is approximately 1, regardless
of the presence of a plateau. 

{\bf Additional checks.}
In the analysis presented in this section, we considered all flares obtained from both the window timing (WT) mode and the photon
counting (PC) mode of the \textit{Swift}-XRT instrument, irrespective of the burst duration $T_{90}$. Therefore, the sample of
flares might be contaminated by episodes of the prompt emission, namely the flares identified at early times
in our analysis might be due to late prompt phase activity. To understand if these flares impact the results, we applied an additional
criteria to the flare sample, considering only the flares characterized by $t_{pk} > T_{90}$, and repeated the analysis. This cut
resulted in the exclusion of 11 flares for bursts with a plateau phase and 4 flares for bursts without a plateau phase. We found that
the conclusions remain unchanged after applying this additional cut. In particular, we find that $\langle \log_{10} t_{pk} \rangle = 2.73 \pm 0.09$
for GRBs with a plateau, compared with $\left< \log_{10} t_{pk}\right> = 2.86 \pm 0.12$ for bursts without a plateau. 

In addition, it has been suggested that there may be a distinction between early \citep[$t_{pk} < 1000{\rm ~s}$, as studied,
{e.g.} by][]{Chincarini+10, Duque+22} and late flares \citep[$t_{pk} \geq 1000{\rm ~s}$, \textit{e.g.}][]{Bernardini+11}.
Considering this possibility as well, we limited the sample to flares for which $T_{90} < t_{pk} < 1000 \ \rm s$,
ensuring that the flares are not contaminated by the prompt emission and occur at early times. As a result of this additional
criteria, the sample size was significantly reduced, comprising 38 flares for GRBs with a plateau phase and 18 flares for GRBs
without a plateau phase. Despite this stringent selection criteria and substantially reduced-sized sample, the ratio $w/t_{\rm pk}$
remains the same, enlightening the solidity of this result.

\subsection{Correlations between flare parameters} \label{sec:correlations}

\begin{figure}
\centering
\includegraphics[width=\linewidth]{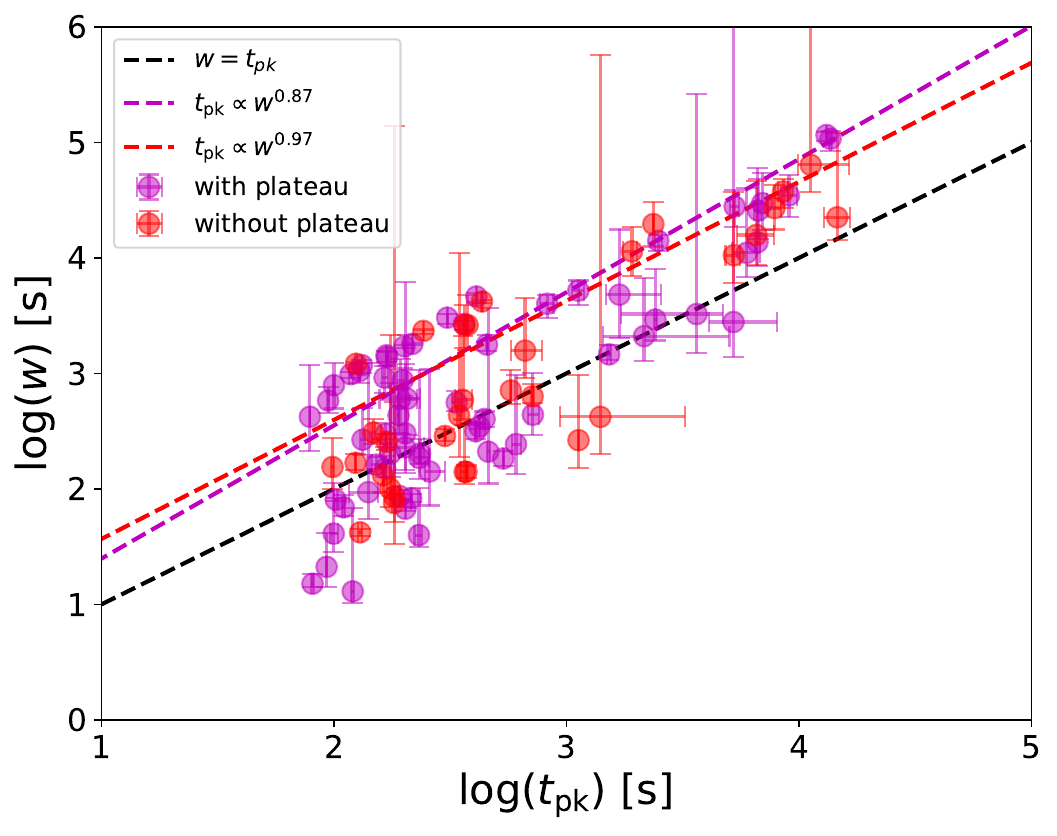}
\caption{The relation between flare peak time ($t_{pk}$) and flare width ($w$). The purple points represent the 42 GRBs (65 flares) with
plateau phases, while the red points represent the 19 GRBs (32 flares) without plateau phases in our sub-samples. The Spearman’s rank
correlation coefficient $r = 0.68~(0.77)$, corresponding to a chance probability of $p \ll 10^{-2} (\ll 10^{-2})$ indicates a strong
monotonic relationship for GRBs with plateau phase (without plateau phase). To model the power-law relationship between $t_{\rm pk}$ and
$w$, we performed a fit and find the power-law index to be $r = 0.87 \pm 4.2 \times 10^{-5}$ and $r = 0.97 \pm 5.9 \times 10^{-5}$ for GRBs
with and  without a plateau phases respectively. The fitted relations are shown as dashed purple and red lines, respectively, matching the
color coding of the data points. Additionally, the line of $w = t_{\rm pk}$ is overlaid.}
\label{fig:scatter_tpk_width}
\end{figure}

The relation between the peak time $t_{\rm pk}$ and the width $w$ of each X-ray flare is presented in Figure \ref{fig:scatter_tpk_width}.
The result of the Spearman’s rank correlation coefficient $r = 0.68~(0.77)$ with a corresponding chance probability of $p \ll 10^{-2} (\ll 10^{-2})$
shows a strong positive correlation between the width $w$ and the $t_{\rm pk}$ for both sub-samples with and without plateau respectively. This
indicates that the longer the peak time $t_{\rm pk}$, the wider the flare. We performed a fit to the correlation using the functional form
$t_{\rm pk} = C \cdot w^r$, where $C$ is the proportionality constant and $r$ is a power-law index. To account for uncertainties, large errors
in the data were capped at the median uncertainty value, reducing the impact of highly uncertain points. We found the power-law index
$r = 0.87 \pm 4.2 \times 10^{-5}$ for burst with plateau and  $r = 0.97 \pm 5.9 \times 10^{-5}$ for bursts without a plateau.

This result is in agreement with the previous findings of \citet{Chincarini+07, Chincarini+10, Bernardini+11, Yi+16}, albeit in our analysis,
we find that this correlation remains true for GRBs with and without a plateau, underlying once more that flares do not seem to be associated
to the phenomena giving rise to the plateau. We note that \citet{Chincarini+07} and \citet{Chincarini+10} interpreted this tight correlation
between $t_{\rm pk}$ and width $w$ as an indication that flares are somewhat different from prompt phase episodes, as they do not follow such
a correlation.

\begin{figure}
\centering
\includegraphics[width=\linewidth]{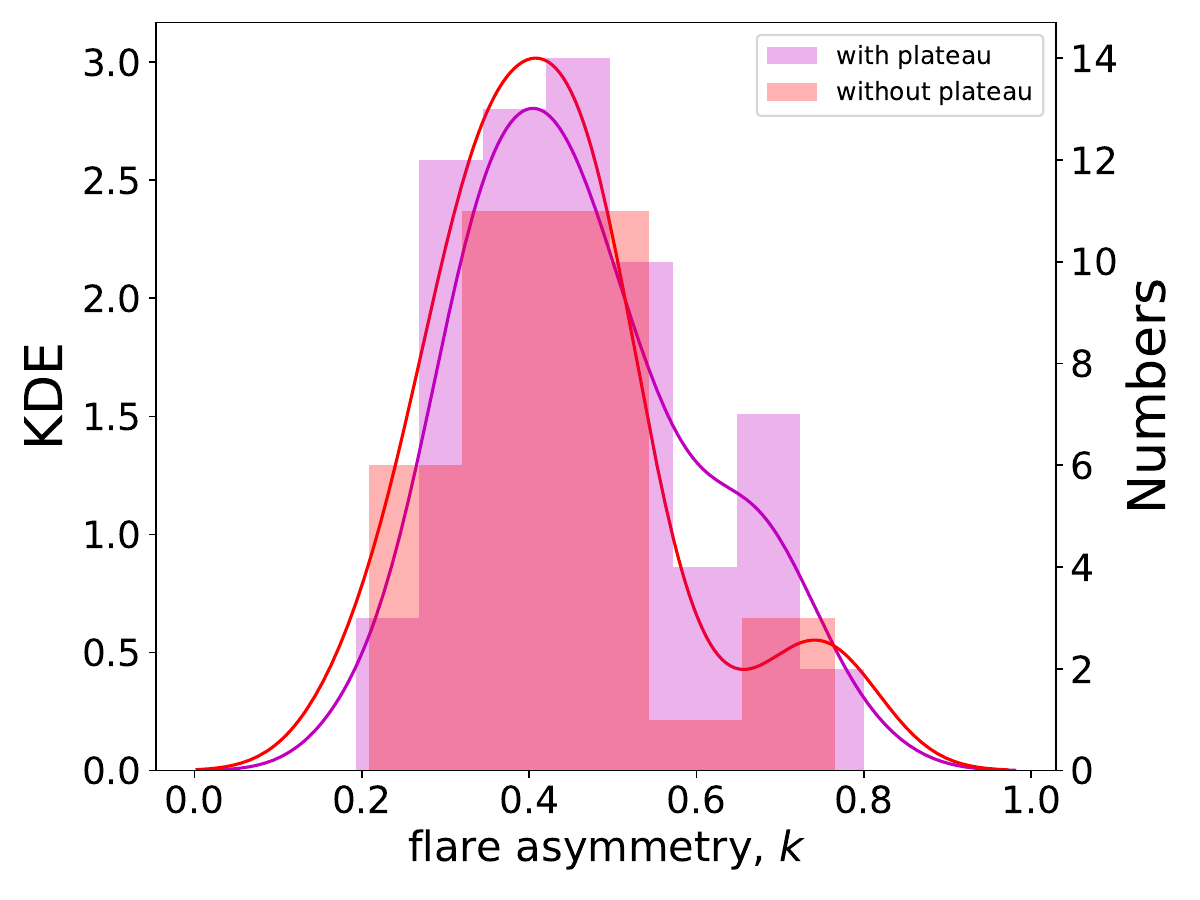}
\caption{Distributions of the flare asymmetry, k. The purple bars represent the 42 GRBs (65 flares) with plateau phases,
while the red bars represent the 19 GRBs (32 flares) without plateau phases. In each panel,
the right-hand ordinate shows the number of burst in each histogram bin and the left-hand ordinate shows the value of
the KDE, which is represented by the purple and red curves for each sample
respectively. We also performed Kolmogorov-Smirnov test for the
flare asymmetry, k (KS test: D = 0.21 and p = 0.23).} 
\label{fig:flare_asymmetry}
\end{figure}

One of the characterizing properties of a flare temporal behavior is its asymmetry. There are two ways of displaying
the flare asymmetry, either by studying $k$ given in Equation \ref{eq:flare_asymetry} or by analyzing the relation between
$t_{\rm rise}$ and $t_{\rm decay}$. In this work, we focus on the first option. In Figure \ref{fig:flare_asymmetry},
we find that the means of $k$\ are $0.46$ and $0.43$ for both the sub-samples with a standard deviation of $0.02$. These values
are consistent with those found in \cite{NBK05, Chincarini+10}, where the median values are $0.49$ and $0.35$ with standard
deviations of $0.26$ and $0.2$ for flares and prompt pulses, respectively.

\section{Discussion: how to define a flare ?} \label{sec:discussion}

In our analysis, to identify a flare over the underlying continuum, we defined a minimum flare amplitude of 0.2 (corresponding to a factor 1.58 in linear scale), which results in a flux cut. This is clearly visible in the
flare-flux variability distribution, $\Delta F_{\rm flare}/F_{\rm cont.}$, shown in Figure \ref{fig:hist_time_flux_varabilities} (right). 
This approach naturally integrates the criteria for significant flare detection into the fitting process without the need for additional
checks, providing a robust criterion for distinguishing significant flares from minor variations in the afterglow. While there is no universally
accepted threshold in the literature, similar empirical criteria have been used in previous studies to identify significant flares. For instance,
\citet{Chincarini+10} employed a combination of visual inspection and statistical analysis to identify flares in X-ray light curves, emphasizing
the importance of clear deviations from the afterglow model. Similarly, \citet{MBB11} utilized timing, duration, and flux increase criteria without
specifying a strict numerical threshold, instead focusing on the statistical significance of the observed deviations. Our chosen threshold is
consistent with these approaches, providing a balance between sensitivity to genuine flares and robustness against stochastic variations.
By ensuring that detected flares exhibit a substantial increase in flux, our method aligns with the practices of \citet{Falcone+07}, who required
statistically significant increases in flux over the underlying afterglow. This threshold, therefore, offers a practical and validated approach to
identifying significant flares in the context of our data and analysis framework.

Furthermore, in our analysis, we fit the data on a logarithmic scale, as opposed to the linear scale used in previous works. This approach provides
several advantages. For instance, we identify one or two smoother flares in each light curve, compared to previous studies that reported between one
and seven or even more flares \citep{Falcone+07, Chincarini+10}. The resulting widths of the flares, as shown in Figure
\ref{fig:hist_peak-time_width_Eiso-flare} (middle), appear broader than those previously reported by \citet{Chincarini+10}, such as in GRB 051117A
(see Fig. 1 therein). Additionally, in Figure \ref{fig:hist_time_flux_varabilities} (right), the flux variation $\Delta F_{\rm flare}/F_{\rm cont.}$
peaks at approximately 5 irrespective of the environment, with a tail extending beyond 100. This result is consistent with the findings of
\citet{Chincarini+10} (see their Fig. 13) and \citet{MBB11} (see their Fig. 7). The only difference is that the tail is shorter, not extending
to 1000, and the second population is less clearly visible than in \citet{MBB11}. This discrepancy might be due to our use of the logarithmic scaling.

Moreover, we calculate the ratio of the flare width to the flare peak time $w/t_{\rm pk}$ based on fits using an asymmetric model \citep{NBK05},
following the approach of \citet{Chincarini+10}. In this model, the width $w$ is defined as the distance between two points where the function has
dropped to $37\%$ (or $1/e$) of its peak value. We find that the ratios of  $w/t_{\rm pk}$ are approximately $0.25\pm0.07$ ($1.78$ linear scale) and $0.22\pm0.09$ ($1.66$ linear scale) for GRBs with and without a plateau,
respectively (see section~\ref{sec:flare}). We find that our values are significantly larger than the $0.23 \pm 0.14$ reported by \citet{Chincarini+10}, which
was comparable to the $0.13 \pm 0.10$ result in \citet{Chincarini+07}, where a narrower Gaussian fit was used under the assumption of symmetric flares\footnote{In
their analysis, the ratio of the flare width to the flare peak time was defined as $\Delta_t/t$ where $\Delta_t$ is the Gaussian width defined by $f=0.05$, which
refers to a specific point in time where the flux of a flare has decreased to $5\%$ of its maximum value, and $t$ is the Gaussian peak \citep{Chincarini+07}.}.
After converting to the Gaussian width at $37\%$, they found the width to be $2.83$ times the Gaussian standard deviation (resulting in $w/t_{\rm pk} = 0.28$).
However, this is still lower than the values we obtain in our analysis. 

In contrast, our result is consistent with what \citet{LP07} found. In their analysis, by assuming a spectral index $\beta \simeq 1$ they found $w/t_{\text{pk}} \sim 0.83$. This value aligns more closely with our findings of approximately $1.78$ and $1.66$ for GRBs with and without
a plateau, respectively, although it still contrasts with the lower value of $0.23 \pm 0.14$ reported by \citet{Chincarini+10}.

\section{Physical Interpretation} \label{physical_interpretation}

By studying GRBs with flares and separating them into two sub-samples, namely with or without a plateau phase, we found that
the properties of the flares are independent of the existence of a plateau phase. This fact offers critical insights into the
proposed emission models underlying the plateau phase. 

\renewcommand{\labelenumi}{(\theenumi)}  
The two key results we find are: 
\begin{enumerate}
  \item The flare peak time, $t_{\rm pk}$, as seen in the Figure \ref{fig:hist_peak-time_width_Eiso-flare} (left), is on average the same for both sub-samples, with and without plateau phases.
  \item Similarly, the ratio of the flare width to the flare peak time, as shown in Figure \ref{fig:hist_time_flux_varabilities} (left), is approximately unity,
irrespective of the existence of plateau. As discussed in section \ref{sec:flare}, removing possible prompt contamination or considering only early flares,
do not change these results. 
\end{enumerate}

X-ray plateaus are ubiquitous in GRB's X-ray light curves. Approximately $60\%$ of GRBs exhibit this phase
\citep{Evans09, Srinivasaragavan20}. This slower than theoretically expected decline in flux in the early X-ray light curve challenged theoretical
modeling for two decades now. Over the years, several models were suggested to explain this phenomenon. Here, we highlight some of the most discussed ones, and confront
their predictions with the obtained results. 

The first model suggested in the literature considers late-time continuous energy injection \citep{Zhang06, Nousek06, Granot+06}. This model
explains the plateau phase by suggesting that late time central engine activity provides energy that is injected into the decelerating
external shock. This additional energy slows down the deceleration of the shock, hence the decay of the light curve. The plateau ends when
this energy injection ceases \citep{Zhang06, Granot+06, GK06, FP+06, Metzger11}. This model successfully predicts achromatic breaks across X-ray and optical
bands for certain bursts, such as GRB 060729 \citep{Grupe+07}. However, it struggles to account for chromatic afterglows, where X-ray and optical light curves
do not show simultaneous breaks \citep{Nousek06}. Extensions of this model, such as the introduction of two-component jets \citep{Racusin+08} or reverse shock
contributions \citep{Uhm+07, Genet+07}, offer solutions but introduce additional complexities and parameters, making it challenging to fully interpret the
variability observed in GRB behaviors \citep{PK01, PK02, Yost+03}.

Despite its strengths, this model predictions conflict with our key findings. Late time central engine activity may lead to the production of some
late time flares. In this case, one would expect that for GRBs with plateau, the average flare occurrence time may be later than for GRBs without
late energy injection (i.e., without plateau). Alternatively, the average ratio of the flare width to its peak time, ($w/t_{\rm pk}$) is expected
to be lower if the energy injection that leads to the flare occurs at later times, as the flares will have shorter time to develop and spread. Both
these predictions are in contradiction to both our key findings. The fact that $t_{\rm pk}$ is the same for both sub-samples suggests it is not a
delayed central engine activity that produces the flares. The constant ratio ($w/t_{\rm pk} \approx 1$) is a strong
indication of a similar motion, such as internal expansion of shells that is proportional to the radius, and originated at early times (during, or close
to the observed prompt phase).

An alternative idea for explaining the plateau is that of an observer located off the jet axis (viewing angle effect), initially suggested by \citet{EG06}.
As was shown by several authors \citep{Toma06, Eichler14}, an observer that is located slightly off the jet axis will observe an X-ray plateau, provided
that the jet is structured (e.g., has an angle-dependent Lorentz factor).  As the jet decelerates, progressively brighter material closer to the
core becomes visible, leading to the shallow flux evolution observed during the plateau phase \citep{BN19, BDD+20a}. This model successfully predicts
the plateau duration and flux as functions of the jet structure and the observer’s viewing angle, establishing correlations between these properties
and the GRB's prompt emission characteristics \citep{BDD+20a}. However, while the model can account for achromatic plateaus in some bursts, it struggles
to explain chromatic afterglows \citep{EG06, OAB+20}.

This model was recently extended to explain X-ray flares as delayed, deboosted versions of gamma-ray pulses produced in the jet’s core, while the plateau
phase is interpreted as deboosted afterglow emission from the core \citep{BGG20b}. In this scenario, energy dissipation that lead to flares would occur,
on the average, at the same radius, $R$ for different GRBs. Different observers will observe the flares at different times, since the observed time is
given by $t \sim R/\mathscr{D}^2 c$, where $\mathscr{D}$ is the Doppler boost, which is angle-dependent. Therefore, it is expected that GRBs with plateaus
would show flares occurring at later times. However, this expectation is contradicted by our finding in point (1), where the flare peak time, $t_{\rm pk}$,
is on average the same for GRBs with and without a plateau. Additionally, \citet{Duque+22} discuss that the off-axis interpretation can only account for
a subset of observed flares, particularly those seen earlier, at $\lesssim 1000$~seconds. However, as discussed in section \ref{sec:flare}, considering early
flares did not affect the result in point (1). Moreover, the consistent patterns and similarity in flare characteristics, such as width, flare asymmetry,
and variability relative to the continuum, suggest that viewing angles might not significantly influence the observed properties.

A third model that was recently suggested is that the plateau originates from emission that occurs during the coasting phase of the propagating forward
shock \citep{ShM12, DPR22}. This phase precedes the self-similar decaying phase that produces the late time afterglow. As was shown by  \citet{DPR22}, this
model can naturally account for both the X-ray and optical signals without requiring complex modifications or additional parameters beyond the standard GRB
"fireball" model framework. Its key difference than the classical GRB "fireball" evolution, is the assumption that during the coasting phase, the GRB
Lorentz factor does not exceeds a few tens (rather than the common assumption of Lorentz factor of a few hundreds). Furthermore, this model requires that
the explosion occurs into a low-density stellar wind environment, as is indeed expected for massive star GRB progenitors. This model effectively explains
the plateau phase as due to synchrotron radiation from particles accelerated by the forward shock during the coasting phase.

This model provides several testable predictions. These include: (1) the expectation that bursts with long plateau phases should
not exhibit high-energy (GeV) emission or strong thermal components; \footnote{GRBs classified as class II in \citet{DPR22} have long and flat plateaus and do not show GeV emissions. Additionally, their variability time $\Delta t_{\min}$ is expected to be $>5$. For example, GRBs classified in class III may have a short and decaying plateau and might exhibit GeV emission.}
(2) both chromatic and a-chromatic breaks are expected, resulting from the fact that
different observed frequencies may be below or above the characteristic synchrotron cooling break \citep[see][for details]{DPR22}.

The results found here are consistent with the predictions of this model. For GRBs with a lower Lorentz factor, energy dissipation processes producing the
flares, such as, e.g., collisions, are expected to occur at smaller radii compared to GRBs with high-Lorentz factor. For example, internal collisions originating
from internal variability of typical time $\delta t$ are expected at radius $r \sim \Gamma^2 c \delta t$, where $\Gamma$ is the jet Lorentz factor. However,
since the observed time is $t \sim r/\Gamma^2 c$, the dependence on the unknown Lorentz factor cancels, and the observed flaring time is similar in both
low- and high- Lorentz factor GRBs. Similar argument holds for the ratio of flare widths to peak time.

\section{Summary and Conclusion} \label{sec:conclusion}
In this paper, we considered the X-ray light curves of 89 GRBs. We find that 61 (69\%) of all GRBs have flares, and 57 (64\%) have
plateaus. However, no correlation was found between the existence of flares and existence of a plateau. We therefore conclude that the existence of flares
is independent of the existence of an X-ray  plateau.

We then analyzed the properties of flares that were detected in those 61 GRBs. Of those GRBs, 42 (68\%) have a plateau, while 19 do not. We found
no statistical difference between the flare peak times and the ratio of flare width to peak time, $w/t_{pk} \sim 1$ between GRBs with and without
plateau. From these results, presented in section \ref{sec:flare}, we conclude that the flare properties of GRBs are similar regardless of the presence
or absence of plateau phases in the GRBs X-ray light curves. 

We then confronted these results with three leading theories discussed in the literature as a possible origin of the plateau: (i) late time energy
injection; (ii) observers located at different viewing angles; and (iii) emission during the coasting phase, which requires the terminal jet Lorentz
factor to not exceed a few tens. We find that the former two models predictions are inconsistent with the observed results. In a late energy injection
model, one expects that at least some of the flares will occur at later times, and may be narrower than without it, which is not seen.
Similarly, for an observer located off the jet axis, the observed flare time should appear later than for an observer located on the line of sight,
due to the different Doppler boosting. The predictions of the low Lorentz factor model, on the other hand, are consistent with the results presented
here. For GRBs with low Lorentz factors, one expects the flares to originate at smaller radius relative to GRBs with larger Lorentz factor. But since
the observed time depends on the Lorentz factor, an observer will see the flares at (average) similar times, regardless of the difference Lorentz factors. 
Our results therefore provide an independent support to the idea that the origin of the plateau phase is GRBs with lower Lorentz factor, of the order of
few tens, as presented in \cite{DPR22}.

This study, employing the approach to fitting the underlying afterglow with physically motivated models and clearly defining the flare
properties, can be expanded to a larger sample in X-rays and extended using data obtained at different wavelengths, such as UV and optical, to address the
debate between X-ray flares and optical flashes. Notably, the recent \textit{SVOM} and upcoming \textit{ULTRASAT} space telescopes aim to increase the number
of GRBs detected in these bands by providing rapid, arcsecond-level localization of bursts to trigger ground-based facilities. In addition, combining
data from Swift and SVOM could allow for a better sampling of the light-curve, in particular between the time range $10^3$ and $10^4$s.

If correct, this model has a strong potential to provide valuable clues about the underlying physics of GRB progenitors and jet dynamics. Coasting
Lorentz factor of several tens that may be in a substantial fraction of GRBs could significantly release physical constraints on the jet acceleration,
and  progenitor properties. It is therefore of high importance to find independent measures that could validate or invalidate this idea. The results
presented here suggest such a strong, independent support.

\begin{acknowledgments}
We wish to thank Dr. Paz Beniamini and Miss. Gowri A. for enlightening discussions on the manuscript.
This work made use of data supplied by the UK Swift Science Data Centre at the University of Leicester. AP, acknowledges support from the European Union  via ERC consolidating grant $\sharp$773062 (acronym O.M.J.) and from the Israel Space Agency via grant number 6766. FR acknowledges support from the Swedish National Space Agency (2021-00180 and 2022-00205)
\end{acknowledgments}

\vspace{5mm}
\facilities{\textit{Swift}-XRT}
\software{MultiNest}

\bibliography{biblio_new}{}
\bibliographystyle{aasjournal}

\appendix
\restartappendixnumbering  

\section{Models used for fitting the data} \label{app:model_definition}

In our analysis, we
consider the following models for fitting the underlying afterglow emission. First, we consider two
classes of models, referring to the medium into which the blast-wave
expands: (i) a "wind" model, in which the ambient density drops with radius as $n(r) \propto r^{-2}$,  and the "interstellar medium (ISM)" model, which assumes a constant ambient density profile, $n(r) \propto r^0$. These two different assumptions affect the evolution of the blast wave Lorentz factor, and are therefore considered separately. We further assume that electrons are accelerated to a power-law in the propagating blast wave, with a power law index $p$. The observed signal is due to synchrotron emission from these electrons. The spectral and temporal slopes depend on the observed frequency, relative to the characteristic synchrotron frequencies: the peak frequency, $\nu_m$ and the cooling frequency, $\nu_c$. E.g., different slopes are expected for $\{ \nu_m, \nu_c \} < \nu^{\rm obs.}$, $\nu_m < \nu^{\rm obs.} < \nu_c$, etc. 
We refer to the
nomenclature defined in \citet{DPR22} to label the different models. Since we are
focusing on data in the X-ray band obtained by the \textit{Swift}-XRT instrument, our analysis
considers the scenarios labeled as C/F and E in \citet{DPR22}, for which the time evolution of the spectrum is given by:
\begin{equation}
    {\rm Case~C/F:} ~~~~~   ~~~~~ \{ \nu_{\rm m}, \nu_{\rm c} \} < \nu^{\rm obs.}  ~~~~~   ~~~~~ F_\nu \propto \left \{ \begin{aligned}
        & t_{\rm obs.}^{(2-p)/2} \nu^{-p/2} & &t_{\rm obs.} < T_{\rm a}      \label{equation:Case_C/F}\\
        & t_{\rm obs.}^{(2-3p)/4} \nu^{-p/2} & &t_{\rm obs.} > T_{\rm a}
    \end{aligned} \right.
\end{equation}
\begin{equation}
    {\rm Case~E:}   ~~~~~   ~~~~~ \nu_{\rm m} < \nu^{\rm obs.} <  \nu_{\rm c}  ~~~~~   ~~~~~ F_\nu \propto \left \{ \begin{aligned}
        & t_{\rm obs.}^{(1-p)/2} \nu^{-(p-1)/2} & &t_{\rm obs.} < T_{\rm a}  \label{equation:Case_E}\\
        & t_{\rm obs}^{(1-3p)/4} \nu^{-(p-1)/2} & &t_{\rm obs.} > T_{\rm a}
    \end{aligned} \right.
\end{equation}
Here, $t_{\rm obs.}$ is the observed time and $T_{a}$ is the end of the plateau time. According to our interpretation, this time marks 
the transition between the coasting and the decelerating phase of the expansion. In this transition, the index of the
electron injection function $p$ is assumed to be constant, thereby linking the 
temporal slope before the plateau to that after the plateau, enforcing rigidity in 
our models.

For the ISM model, we refer to the two segments labeled as G and H in \citet{SPN98}: 
\begin{equation}
    {\rm Case~G:} ~~~~~   ~~~~~ \quad \nu_{\rm m} < \nu^{\rm obs.} < \nu_{\rm c} ~~~~~   ~~~~~ F_\nu \propto t_{\rm obs.}^{(3-3p)/4} \nu^{-(p-1)/2} 
    \label{equation:Case_G}
\end{equation}
\begin{equation}
    {\rm Case~ H:} ~~~~~   ~~~~~ \quad \{ \nu_{\rm m}, \nu_{\rm c} \} < \nu^{\rm obs.} ~~~~~   ~~~~~  F_\nu \propto t_{\rm obs.}^{(2-3p)/4} \nu^{-p/2} 
    \label{equation:Case_H}
\end{equation}
In this case, we assume that the transition between coasting
phase and self-similar phase happens before the first \textit{Swift}-XRT observations or during the steep decay.
Consequently, in our analysis, we do not consider the coasting phase for
ISM models.

The Cases C/F and E for the wind environment and the cases H and G for the ISM environment, represent
different spectral regimes depending on the cooling state of the plasma behind the shock, namely on
the position of the cooling frequency $\nu_{\rm c}$ with respect to the injection frequency
$\nu_{\rm m}$. Finally, on top of each of emission model, we further consider the possibility
of observing (i) a steep decay, assumed in this analysis to be associated to the end of the prompt phase,
(ii) a jet break and (iii) zero, one or two  flares. This led us to fit a total of 36 models to each observed
afterglow light curve. 12 models are considered for the "wind" environment scenario, differ by existence/in-existence of jet break, number of flares (0,1,2) and spectral regime (C/F or E). 24 models are considered for the "ISM" case, which, in addition to the above, consider existence or in-existence of a steep decay (in the "wind" model, which is used to fit GRBs with plateau, a steep decay always assumed to exist). 
The models are summarized in Table \ref{tab:model_comparison}.

All afterglow components at the exception of the flares are assumed to be power-law
functions of time, namely the flux in the X-ray band is
$F_x(t) \propto t^a$. The power-law index $a$ is motivated by the considered component
and the physical scenario as described in the equations above, at the exception of the index of the fast decay
which is constrained to be greater than $2$. 
All components making an empirical
afterglow model form a continuous function in time. The analysis is performed
in log space, meaning that the variable becomes 
$T \equiv \log(t)$ and the observable is $\mathfrak{F} \equiv \log(F)$. For instance,
a model of afterglow in the wind environment with a steep decay, a jet break
and such that $\{ \nu_{\rm m}, \nu_{\rm c} \} \ll \nu^{\rm obs.}$, is expressed as
\begin{equation}
    \mathfrak{F} = \left \{\begin{aligned}
& \alpha T + A  & ~~~~~~~~ & T < T_1 \\
& \beta T  + A + (\alpha - \beta) T_1  & ~~~~~~~~ & T_1 < T < T_2 \\
& \zeta T  + A + (\alpha - \beta) T_1 + (\beta - \zeta) T_2  & ~~~~~~~~ & T_2 < T < T_3 \\
& (\zeta + q) T  + A + (\alpha - \beta) T_1 + (\beta - \zeta) T_2 + ( \zeta - \eta ) T_3   & ~~~~~~~~ & T_3 > T 
    \end{aligned}
    \right.
\label{equation:fitting}
\end{equation}
Here, $A$ represents the normalization of the flux, $T_1$ is the transition
time between the steep decay and the first afterglow segment, namely
the plateau, $T_2\equiv T_{\rm a}$ is the time at which the blast-wave transitions from the plateau to a steeper decay (interpreted as 
coasting to the decelerating phase), and $T_3$ is the time
of the jet break. For this specific regime, the parameters $\beta$ and $\zeta$
are not independent: they both depends on the power-law index $p$ of the injected 
electrons at the shock front. We use this constraint, by 
setting $\beta $ and $\zeta $ to the parameter dependencies presented in Equations
\ref{equation:Case_C/F}, \ref{equation:Case_E}, \ref{equation:Case_G},
\ref{equation:Case_H}, and consider $p$ to be the free model parameter.
For the example of Equation \ref{equation:fitting}, we have $\beta = (2-p)/2$
and $\zeta = (2-3p)/4$. In addition, after the jet break, the afterglow slope is
steeper than $\zeta$ and rather than fitting directly for the slope, we fit for
the difference $q$ between the post-jet-break slope and $\zeta$, such that the new
slope is $\zeta + q$. We allow for the extra degree of freedom in the choice of the value of $q$, as the jet break is not sharp, but rather is observed over a long duration, in particular for observers that are off the jet axis. These techniques add some rigidity to the models and
allow for a straightforward interpretation of the results. An example of the fitting results are presented in Figure \ref{fig:fitting_example} in Appendix \ref{app:fitting_example_190719C}. 

\begin{table}[H]
\addtocounter{table}{-1}
\centering
\caption{Comparison of models for different theoretical scenarios. For wind scenarios (Cases C/F and E), the models include conditions for steep decay, jet break, and number of flares. All models incorporate the early steep decay phase, with optional jet break and flare presence. For ISM scenarios (Cases H and G), the models include similar conditions with additional differentiation, where the early steep decay is also optional. The table details the presence of steep decay, jet break, and number of flares for each model.}
\begin{tabular}{c|c|c|c|c}
\toprule
Environment & \textbf{Model Name} & \textbf{Steep Decay} & \textbf{Jet Break} & \textbf{Number of Flares} \\
\midrule
Wind & \textbf{Case C/F and E} \\
& C1 / E1 & Yes & No  & 0 \\
& C2 / E2 & Yes & Yes & 0 \\
& C3 / E3 & Yes & No  & 1 \\
& C4 / E4 & Yes & No  & 2 \\
& C5 / E5 & Yes & Yes & 1 \\
& C6 / E6 & Yes & Yes & 2 \\
\hline
ISM & \textbf{Case H and G} \\
 & H1 / G1 & No & No  & 0 \\
 & H2 / G2 & No & No & 1 \\
 & H3 / G3 & No & No  & 2 \\
 & H4 / G4 & Yes & No  & 0 \\
 & H5 / G5 & Yes & No & 1 \\
 & H6 / G6 & Yes & No & 2 \\
 & H7 / G7 & No & Yes  & 0 \\
 & H8 / G8 & No & Yes & 1 \\
 & H9 / G9 & No & Yes  & 2 \\
 & H10 / G10 & Yes & Yes  & 0 \\
 & H11 / G11 & Yes & Yes & 1 \\
 & H12 / G12 & Yes & Yes & 2 \\
\bottomrule
\end{tabular}
\label{tab:model_comparison}
\end{table}


\subsection{Example fit to the X-ray Light curve of GRB 190719C} \label{app:fitting_example_190719C}

\begin{figure}[H]
    \centering
    \includegraphics[width=0.45\linewidth]{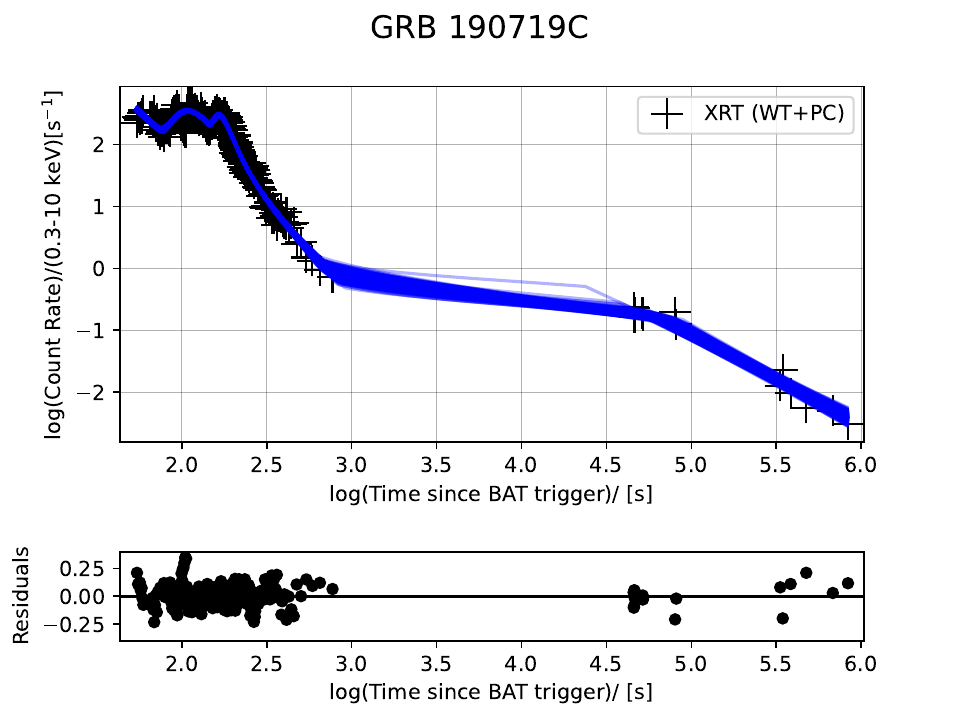}
    \includegraphics[width=0.45\linewidth]{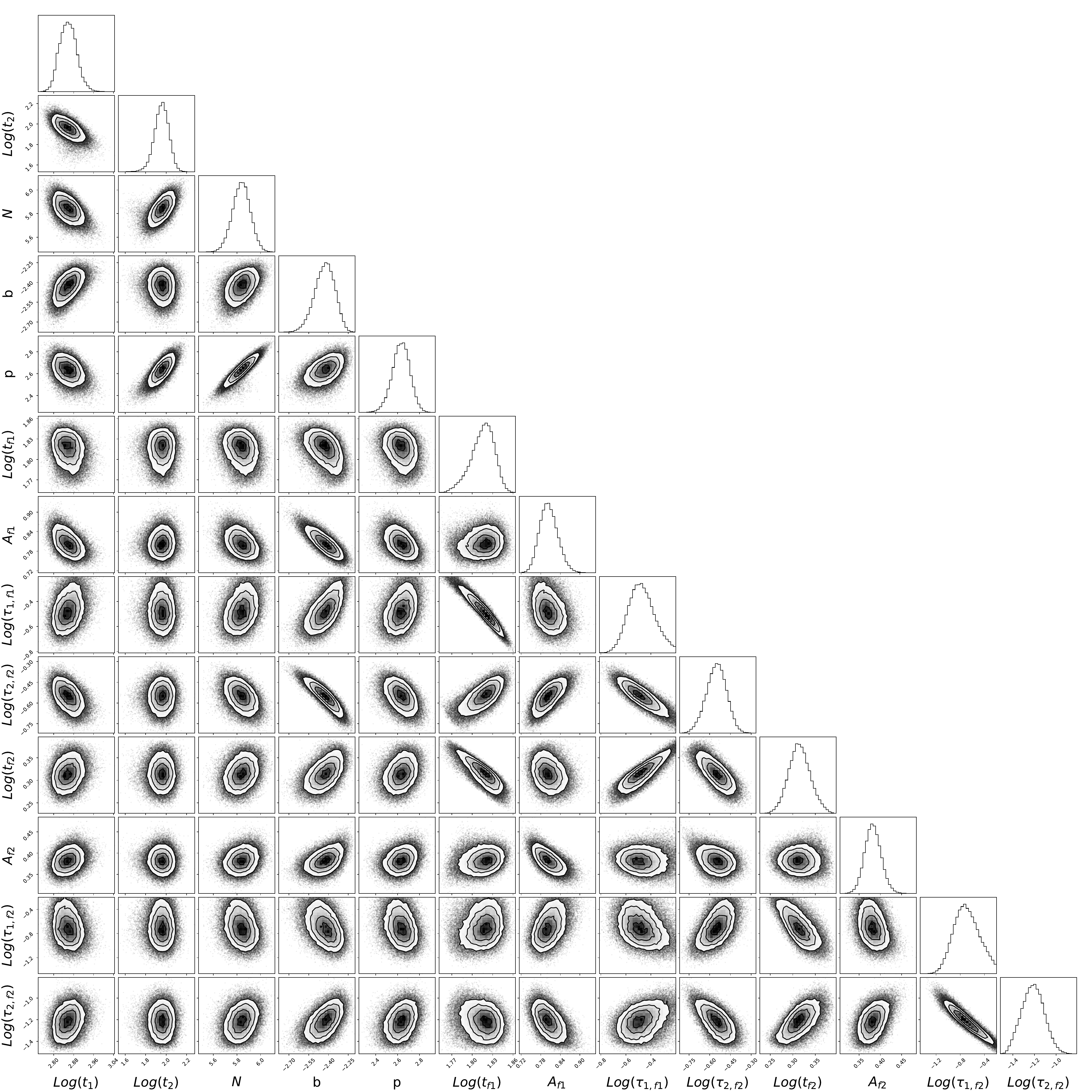}
    \caption{Left-top: The X-ray LC of GRB 190719C. The black crosses represent the XRT-WTSLEW, XRT-WT, XRT-PC mode data. The errors correspond to a significance of one sigma. The blue lines represent the posterior distributions of the best fit model. Left-bottom: The residuals between the best model and the data. Right: The corner plot of the posterior probability distributions of the fit parameters 
and the covariances between the fit parameters.}
    \label{fig:fitting_example}
\end{figure}

To visualize the fitting process, in Figure \ref{fig:fitting_example} (left-top), 
we illustrate the X-ray count light-curve of GRB 190719C obtained by the \textit{Swift}-XRT instrument. The data is fitted in log scale, therefore, for consistency we present 
the data in log-scale as well. 
The data are superposed with the fit results (blue) from a selected best model where three-segments
BPL model, i.e. with two break times, and two Norris functions are used to fit two flares. It represents the case C/F in the wind where steep decay, plateau phase and the late afterglow decay are present, but no jet break. This model is shown as C4 in Table \ref{tab:model_comparison}.
In Figure \ref{fig:fitting_example} (left-bottom), we present the residuals 
between the best model and the data which shows how well the model describes the data.
The corner plot of the posterior probability distributions of the fit parameters 
and the covariances between the fit parameters is displayed in Figure \ref{fig:fitting_example} (right). 

\section{Bursts excluded from the sample} \label{app:exluded_bursts}

Out of the 100 GRBs in our sample, we exclude 11 GRBs listed in Table \ref{tab:list_11GRBs}. The exclusion criteria was misidentification of flares, which compromised the underlying afterglow emission and failures of the different afterglow models to adequately fit the data. When this happened, identified flares extend more than the data, most of the time up to $> 10^5$~s which corresponds to the more than the half of the dynamical range.  As a result, the allowed variability in the flare shape replaces the underlying afterglow, which is at nearly all time dominated by flare emission. Therefore, such a exclusion was done to avoid misinterpretation of the model parameters.

\begin{table}[H]
\addtocounter{table}{-1}
\caption{The list of excluded 11 GRBs from the main sample. Column 1: GRB name, Column 2: The flare width, Column 3: The flare isotropic energy.
\label{tab:list_11GRBs}}
\centering
\small
\begin{tabular}{c|c|c}
\hline
GRB & $w$ & $E_{\rm iso,f}$  \\
 name & (s) & (ergs)  \\
\hline
\hline
221009A & $0.4\times10^5$ & $10^{51}$  \\
210610A & $0.7\times10^7$& $10^{52}$ \\
210104A & $0.5\times10^7$& $10^{51}$ \\
201221A & $0.3\times10^5$ & $10^{52}$ \\
190114C & $0.3\times10^7$ & $10^{51}$ \\
170728B & $0.3\times10^7$ & $10^{52}$  \\
161129A & $0.8\times10^7$& $10^{51}$ \\
160117B & $0.4\times10^8$ & $10^{51}$  \\
150910A & $0.2\times10^7$ &$10^{52}$ \\
150403A & $0.1\times10^8$ & $10^{53}$ \\
140614A & $0.8\times10^8$ & $10^{52}$ \\
\hline
\end{tabular}
\end{table}

\section{Sample of 61 GRBs with flares and derived parameters} \label{app:GRBs_with_flares_derived_params}
In our analysis, the fit parameters are computed using the maximum posterior estimate (MAP) method. The derived parameters and their errors are calculated using the marginalized posterior distributions (MPD). As discussed in section \ref{sec:model_comp}, the selection of the best fit models is based on the AIC, and AICc criteria. The results show that the AICc condition is consistently satisfied by the data. Additionally, in most cases, the AIC condition is also met with the same model. Therefore, we only present the AICc in Table \ref{tab:derived_params1}, along with the derived parameters obtained from the best fit parameters. The parameter errors are computed by using the credible intervals (e.g., 68\% for $1\sigma$) derived directly from the posterior samples of the fit parameters.

\begin{deluxetable}{ccccccccc}[H]
\tablecaption{Derived parameters for selected GRBs with flares (Part 1). Column 1: GRB name, Column 2: Best model, Columns 3: corrected Akaike Information Criterion of the best model, Columns 4-9: Derived parameters with their errors—flare peak time, flare width, flare asymmetry, total flare isotropic energy in units of $10^{50} \rm erg$, flare width to the flare peak time ratio, and flare peak flux-to-underlying continuum ratio. When the best model has two flares, the entry corresponding to this GRB is on two rows.
\label{tab:derived_params1}}
\tablehead{
\colhead{GRB} & \colhead{Best} & \colhead{AICc}& \colhead{$t_{\rm pk}$} & \colhead{$w$} & \colhead{$k$} &  \colhead{$E_{\rm iso,f}$} & \colhead{$w/t_{\rm pk}$} & \colhead{$\Delta F_{\rm flare}/F_{\rm cont.}$} \\
 \colhead{name}  & \colhead{model} & \colhead{} & \colhead{(s)} & \colhead{(s)}  & \colhead{} &  \colhead{($10^{50} \rm erg$)} & \colhead{}  & \colhead{} 
}
\startdata
220117A &   C4 & 251 &  $189^{+1.6}_{-2.04}$ &        $434^{+240}_{-157}$ & $0.37^{+0.03}_{-0.03}$ &         $360^{+22.6}_{-25.5}$ & $2.3^{+1.3}_{-0.8}$ &    $9.7^{+4.2}_{-2.6}$ \\
&   &  &  $1124^{+67}_{-68}$ &     $5203^{+1054}_{-1499}$ & $0.45^{+0.01}_{-0.01}$ &       $373^{+55.1}_{-68.2}$ & $4.6^{+0.8}_{-1.2}$ &    $6.9^{+2.5}_{-1.9}$ \\
220101A &   C5 & 801  & $167^{+6.7}_{-7.8}$ &          $298^{+68}_{-75}$ & $0.50^{+0.02}_{-0.02}$ &           $5769^{+785}_{-115}$ & $1.8^{+0.3}_{-0.4}$ &   $11.2^{+4.4}_{-3.6}$ \\
211024B &   C4 &  442   & $6621^{+169}_{-186}$ &   ${1.4\times10^4}^{+2\times10^4}_{-0.6\times10^4}$ & $0.38^{+0.05}_{-0.05}$ &       $123^{+15.4}_{-31.1}$ & $2.1^{+3.0}_{-1.0}$ &      $178^{+52}_{-77}$ \\
 &   &   &  ${1.4\times10^4}_{-153}^{+145}$ & ${1.1\times10^5}^{+1.6\times10^4}_{-2.6\times10^4}$ & $0.32^{+0.01}_{-0.01}$ &     $103^{+3.5}_{-3.5}$ & $7.9^{+1.2}_{-1.9}$ & $36.5^{+13.5}_{-10.5}$ \\
210905A &  G3 & 272  & $98^{+1.1}_{-1.2}$ &          $155^{+91}_{-69}$ & $0.37^{+0.04}_{-0.03}$ &        $239^{+18.6}_{-21.5}$ & $1.6^{+0.9}_{-0.7}$ &    $2.3^{+0.3}_{-0.2}$ \\
 &   &  & $371^{+1.4}_{-1.2}$ &      $140^{+17.9}_{-16.1}$ & $0.75^{+0.01}_{-0.02}$ &      $1208^{+101}_{-91}$ & $0.4^{+0.05}_{-0.04}$ &   $17.4^{+1.0}_{-0.9}$ \\
210722A &  G9 & 278  &   $124^{+0.8}_{-0.8}$ &        $1210^{+54}_{-102}$ & $0.30^{+0.004}_{-0.004}$ &      $11.9^{+0.5}_{-0.5}$ & $9.7^{+0.4}_{-0.8}$ &    $2.3^{+0.1}_{-0.1}$ \\
&   &   & $8545^{+439}_{-468}$ &   $3.8\times{10^{4}}^{+1\times10^4}_{-1.2\times10^4}$ & $0.45^{+0.02}_{-0.01}$ &        $11.7^{+1.1}_{-1.2}$ & $4.4^{+1.0}_{-1.3}$ &    $3.3^{+0.5}_{-0.4}$ \\
210702A &  H8 & 523 &  $5245^{+308}_{-435}$ &   $1\times{10^{4}}^{+1.3\times10^4}_{-0.6\times10^4}$ & $0.47^{+0.09}_{-0.05}$ &         $37.9^{+13.3}_{-10.9}$ & $2.0^{+2.4}_{-1.1}$ &    $1.0^{+0.2}_{-0.2}$ \\
210610B &   C5 &  651 &  $102^{+1.7}_{-1.9}$ &         $80^{+2.7}_{-2.4}$ & $0.57^{+0.02}_{-0.02}$ &            $46.9^{+1.3}_{-0.9}$ & $0.8^{+0.03}_{-0.02}$ &   $37.6^{+5.3}_{-3.9}$ \\
210504A &  E3 & 104 &  $460^{+6.9}_{-7.0}$ &        $210^{+601}_{-134}$ & $0.41^{+0.12}_{-0.10}$ &           $11.8^{+1.9}_{-2.9}$ & $0.5^{+1.3}_{-0.3}$ &    $4.6^{+0.8}_{-0.7}$ \\
210420B &  E3 & 121 & $716^{+42.9}_{-33.6}$ &        $438^{+362}_{-180}$ & $0.64^{+0.09}_{-0.12}$ &          $5.1^{+1.6}_{-1.4}$ & $0.6^{+0.5}_{-0.2}$ &    $2.8^{+0.4}_{-0.3}$ \\
210411C &  G9 &  70 & $174^{+3.9}_{-4.7}$ &       $101^{+308}_{-37.7}$ & $0.48^{+0.16}_{-0.12}$ &         $32.9^{+12.1}_{-8.1}$ & $0.6^{+1.7}_{-0.2}$ &    $4.9^{+0.7}_{-0.6}$ \\
 &   &   &   $663^{+113}_{-92}$ &      $1573^{+1650}_{-865}$ & $0.47^{+0.06}_{-0.04}$ &          $30.5^{+11.6}_{-10.8}$ & $2.4^{+2.1}_{-1.2}$ &    $0.9^{+0.2}_{-0.2}$ \\
201104B &   C3 & 112  & $202^{+3.2}_{-3.6}$ &        $303^{+913}_{-242}$ & $0.30^{+0.12}_{-0.06}$ &           $3.1^{+0.65}_{-0.68}$ & $1.5^{+4.5}_{-1.2}$ &    $0.7^{+0.1}_{-0.1}$ \\
200205B &   C6 & 523 &   $455^{+1.9}_{-1.7}$ &       $1774^{+348}_{-221}$ & $0.31^{+0.01}_{-0.01}$ &        $168^{+2.3}_{-2.2}$ & $3.9^{+0.8}_{-0.5}$ &   $17.8^{+1.0}_{-0.9}$ \\
&   &  &   $534^{+2.9}_{-2.9}$ &      $183^{+27.5}_{-25.8}$ & $0.74^{+0.01}_{-0.01}$ &       $47.5^{+2.1}_{-2.4}$ & $0.3^{+0.05}_{-0.05}$ &    $2.3^{+0.2}_{-0.2}$ \\
191221B &   C5 &  954 & $169^{+1.9}_{-2.0}$ &       $1382^{+140}_{-244}$ & $0.33^{+0.01}_{-0.004}$ &          $28.9^{+1.0}_{-0.9}$ & $8.2^{+0.8}_{-1.4}$ &    $2.5^{+0.1}_{-0.1}$ \\
190829A &   C5 & 789 & $1524^{+29.4}_{-26.6}$ &       $1460^{+280}_{-154}$ & $0.54^{+0.03}_{-0.03}$ &         $28.9^{+1.0}_{-0.9}$ & $1.0^{+0.2}_{-0.1}$ &    $3.0^{+0.1}_{-0.1}$ \\
190719C &   C4 & 685 &   $175^{+1.1}_{-1.1}$ &         $202^{+286}_{-96}$ & $0.35^{+0.06}_{-0.06}$ &            $589^{+20}_{-25}$ & $1.2^{+1.6}_{-0.5}$ &    $1.4^{+0.1}_{-0.1}$ \\
 &   & &  $133^{+3.2}_{-3.0}$ &         $266^{+167}_{-80}$ & $0.43^{+0.03}_{-0.04}$ &      $154^{+9.4}_{-9.5}$ & $2.0^{+1.3}_{-0.6}$ &    $5.4^{+0.4}_{-0.4}$ \\
190114A & E6 & 130 &   $232^{+6.5}_{-10.1}$ &      $39.8^{+37.2}_{-9.1}$ & $0.55^{+0.09}_{-0.11}$ &         $67^{+26}_{-22}$ & $0.2^{+0.2}_{-0.04}$ &    $3.0^{+1.6}_{-0.9}$ \\
 &   &  &  $1688^{+700}_{-219}$ &     $4800^{+6242}_{-4158}$ & $0.45^{+0.10}_{-0.06}$ &       $34.2^{+61}_{-25}$ & $2.5^{+2.4}_{-2.1}$ &    $0.9^{+0.8}_{-0.2}$ \\
190106A &   C6 & 290 &    $100^{+0.5}_{-0.5}$ &        $790^{+350}_{-368}$ & $0.22^{+0.02}_{-0.01}$ &          $36.6^{+2.5}_{-2.5}$ & $7.9^{+3.5}_{-3.7}$ &    $1.3^{+0.1}_{-0.1}$ \\
 &    &  &  $258^{+38.9}_{-12.1}$ &         $141^{+289}_{-96}$ & $0.56^{+0.15}_{-0.20}$ &         $1.5^{+1.7}_{-0.61}$ & $0.5^{+0.9}_{-0.3}$ &    $0.7^{+0.2}_{-0.1}$ \\
181110A &   C6  & 588 &    $168^{+1.2}_{-1.2}$ &       $1434^{+121}_{-183}$ & $0.32^{+0.01}_{-0.004}$ &          $798^{+16.4}_{-16.7}$ & $8.5^{+0.7}_{-1.1}$ &    $7.2^{+0.9}_{-0.8}$ \\
 &    &  &   $118^{+2.1}_{-2.1}$ &        $984^{+36.1}_{-67}$ & $0.34^{+0.01}_{-0.004}$ &         $536^{+17.9}_{-18.3}$ & $8.3^{+0.3}_{-0.6}$ &    $6.9^{+0.6}_{-0.5}$ \\
181020A & G12 & 1669  &  $242^{+0.6}_{-0.6}$ &        $2331^{+64}_{-123}$ & $0.31^{+0.002}_{-0.002}$ &           $942^{+10.3}_{-10.4}$ & $9.6^{+0.3}_{-0.5}$ &    $6.2^{+0.1}_{-0.1}$ \\
 &  &   &  $375^{+0.7}_{-0.8}$ &     $2623^{+1565}_{-1311}$ & $0.21^{+0.02}_{-0.01}$ &    $103^{+4.7}_{-4.5}$ & $7.0^{+4.2}_{-3.5}$ &    $1.1^{+0.04}_{-0.04}$ \\
181010A &   C5 & 208 &  $140^{+13.2}_{-27.0}$ &          $94^{+242}_{-50}$ & $0.53^{+0.18}_{-0.16}$ &          $1.3^{+1.0}_{-0.6}$ & $0.8^{+1.5}_{-0.4}$ &    $1.1^{+0.7}_{-0.2}$ \\
180728A &  E4 & 978 & $2411^{+38.1}_{-37.3}$ &     $2927^{+2965}_{-1228}$ & $0.43^{+0.06}_{-0.05}$ &        $0.26^{+0.46}_{-0.07}$ & $1.2^{+1.2}_{-0.5}$ &    $1.4^{+0.1}_{-0.1}$ \\
 &  &  & $5226^{+2268}_{-1252}$ &    $2784^{+18002}_{-1935}$ & $0.44^{+0.18}_{-0.12}$ &      $0.3^{+1.7}_{-0.2}$ & $0.5^{+3.0}_{-0.3}$ &   $1.9^{+18.4}_{-1.2}$ \\
180720B &   C4 & 3434 &    $110^{+1.8}_{-1.7}$ &       $69^{+17.4}_{-12.5}$ & $0.71^{+0.04}_{-0.04}$ &          $32.2^{+3.4}_{-3.2}$ & $0.6^{+0.2}_{-0.1}$ &    $1.2^{+0.1}_{-0.1}$ \\
 &   &  &   $362^{+5.8}_{-5.4}$ &       $2675^{+177}_{-315}$ & $0.36^{+0.01}_{-0.004}$ &        $53.7^{+2.5}_{-2.5}$ & $7.4^{+0.5}_{-0.8}$ &    $1.4^{+0.1}_{-0.1}$ \\
180624A &   E4 & 1487  &   $192^{+1.0}_{-0.7}$ &        $87^{+10.6}_{-6.7}$ & $0.66^{+0.02}_{-0.02}$ &           $309^{+12.1}_{-9.1}$ & $0.5^{+0.1}_{-0.04}$ &   $26.9^{+1.9}_{-1.3}$ \\
&  &   &  $444^{+5.6}_{-7.0}$ &          $403^{+60}_{-65}$ & $0.63^{+0.01}_{-0.02}$ &          $524^{+16.5}_{-22.2}$ & $0.9^{+0.1}_{-0.1}$ &     $53^{+4.2}_{-5.3}$ \\
180620B &  E6 & 367 &   $165^{+5.4}_{-4.9}$ &        $918^{+206}_{-280}$ & $0.40^{+0.02}_{-0.01}$ &          $61.9^{+2.9}_{-3.1}$ & $5.6^{+1.2}_{-1.7}$ &    $3.9^{+0.6}_{-0.5}$ \\
&   &   &  $5956^{+807}_{-841}$ &   $1.1\times{10^{4}}^{+1.4\times10^4}_{-0.5\times10^4}$ & $0.49^{+0.09}_{-0.04}$ &       $8.2^{+4.1}_{-2.8}$ & $1.9^{+2.3}_{-0.8}$ &    $1.1^{+0.2}_{-0.2}$ \\
180329B &  E6 & 440 &    $166^{+1.1}_{-1.1}$ &      $150^{+38.8}_{-25.8}$ & $0.43^{+0.02}_{-0.02}$ &         $161^{+4.5}_{-7.1}$ & $0.9^{+0.2}_{-0.2}$ &     $94^{+6.2}_{-6.9}$ \\
 &   &  &   $234^{+2.3}_{-2.0}$ &        $202^{+58}_{-41.2}$ & $0.39^{+0.02}_{-0.03}$ &       $44^{+4.5}_{-3.7}$ & $0.9^{+0.2}_{-0.2}$ &    $2.3^{+0.5}_{-0.3}$ \\
180325A &   C3 & 381&    $81^{+0.8}_{-0.9}$ &       $15.1^{+3.0}_{-1.1}$ & $0.56^{+0.06}_{-0.05}$ &          $93.8^{+9.5}_{-8.1}$ & $0.2^{+0.03}_{-0.01}$ &   $14.0^{+1.2}_{-1.0}$ \\
180205A &  G8 & 89 &   $181^{+1.8}_{-1.8}$ &        $86^{+161}_{-43.7}$ & $0.42^{+0.07}_{-0.08}$ &        $4.8^{+0.7}_{-10.6}$ & $0.5^{+0.9}_{-0.2}$ &    $4.5^{+1.0}_{-0.7}$ \\
171222A &   C3 & 212 &  $337^{+19.0}_{-17.1}$ &         $558^{+131}_{-94}$ & $0.51^{+0.02}_{-0.03}$ &          $225^{+20.3}_{-20.0}$ & $1.6^{+0.3}_{-0.2}$ &    $4.5^{+0.8}_{-0.8}$ \\
170714A &   C4 & 3227 &  $5285^{+211}_{-258}$ &   $2.8\times{10^{4}}^{+1\times10^4}_{-1.1\times10^4}$ & $0.38^{+0.02}_{-0.01}$ &   $120^{+9.8}_{-9.2}$ & $5.3^{+1.7}_{-2.1}$ &    $4.8^{+0.2}_{-0.2}$ \\
&   &  &   $1.3\times{10^{4}}^{+82}_{-86}$ &  $1.2\times{10^{5}}^{+0.8\times10^4}_{-1.5\times10^4}$ & $0.32^{+0.01}_{-0.004}$ &   $81.8^{+2.2}_{-2.2}$ & $8.8^{+0.6}_{-1.1}$ &    $7.6^{+0.6}_{-0.5}$ \\
170705A &   C5 & 992 &    $217^{+0.7}_{-0.7}$ &     $1800^{+19.0}_{-39.4}$ & $0.35^{+0.001}_{-0.001}$ &       $245^{+2.1}_{-2.1}$ & $8.3^{+0.1}_{-0.2}$ &     $76^{+2.9}_{-2.9}$ \\
\enddata
\end{deluxetable}

\begin{deluxetable}{ccccccccc}
\renewcommand{\thetable}{}
\makeatletter
\renewcommand{\@makecaption}[2]{#2} 
\makeatother
\caption{\textbf{Table \ref{tab:derived_params1}} (Continued) \label{tab:derived_params2}}
\tablehead{
\colhead{GRB} & \colhead{Best} & \colhead{AICc} & \colhead{$t_{\rm pk}$} & \colhead{$w$} & \colhead{$k$} &  \colhead{$E_{\rm iso,f}$} & \colhead{$w/t_{pk}$} & \colhead{$\Delta F_{\rm flare}/F_{\rm cont.}$} \\
 \colhead{name}  & \colhead{model} & \colhead{} & \colhead{(s)} & \colhead{(s)}  & \colhead{} &  \colhead{($10^{50} \rm erg$)} & \colhead{}  & \colhead{} 
}
\startdata
170607A &  E4 & 576 &  $152^{+8.6}_{-7.3}$  &  $162^{+24.2}_{-12.7}$ & $0.58^{+0.03}_{-0.03}$ &         $3.6^{+0.2}_{-0.2}$ & $1.1^{+0.1}_{-0.1}$ &    $1.8^{+0.2}_{-0.1}$ \\
        &     &     &  $9050^{+685}_{-708}$ &  $3.5\times{10^{4}}^{+1.5\times10^4}_{-1.5\times10^4}$ & $0.45^{+0.03}_{-0.02}$ &      $1.9^{+0.4}_{-0.5}$ &  $3.8^{+1.5}_{-1.5}$ &   $0.9^{+0.1}_{-0.1}$ \\
170604A &  3H & 535 &  $130^{+2.4}_{-2.2}$  &  $42.0^{+2.8}_{-3.0}$ & $0.73^{+0.01}_{-0.02}$ &           $37.3^{+3.3}_{-3.1}$ &  $0.3^{+0.02}_{-0.03}$ &   $9.5^{+0.9}_{-0.8}$ \\
        &     &     &  $364^{+1.0}_{-1.1}$  &  $2623^{+1037}_{-1192}$ & $0.28^{+0.02}_{-0.01}$ &           $84^{+1.9}_{-1.8}$ &  $7.2^{+2.8}_{-3.3}$ &  $12.1^{+0.4}_{-0.4}$ \\
170531B &  G3 & 397 &  $168^{+0.4}_{-0.4}$  &  $255^{+40.1}_{-39.7}$ & $0.38^{+0.01}_{-0.01}$ &           $52^{+1.7}_{-2.3}$ &  $126^{+1.3}_{-1.4}$ &  $48.2^{+1.8}_{-2.5}$ \\
        &     &     &  $576^{+0.6}_{-0.6}$  &  $711^{+294}_{-187}$ & $0.33^{+0.02}_{-0.02}$ &             $98^{+5.2}_{-4.3}$ &  $76^{+1.1}_{-1.1}$ &    $62^{+2.0}_{-2.8}$ \\
170519A &  E4 & 820 &  $203^{+0.4}_{-0.4}$  &  $67^{+2.9}_{-2.8}$ & $0.72^{+0.01}_{-0.01}$ &             $28.4^{+0.5}_{-0.5}$ &  $0.3^{+0.01}_{-0.01}$ &  $49.7^{+1.7}_{-1.7}$ \\
        &     &     &  $2153^{+1824}_{-883}$& $2099^{+2432}_{-1028}$ & $0.64^{+0.12}_{-0.16}$ &          $2.1^{+2.0}_{-1.0}$ &  $0.9^{+0.5}_{-0.3}$ &   $2.6^{+5.5}_{-1.2}$ \\
170405A &  H9 & 420 &  $162^{+0.7}_{-0.7}$  &  $131^{+95}_{-44.8}$ & $0.36^{+0.03}_{-0.04}$ &            $242^{+10.3}_{-11.7}$ &  $0.8^{+0.6}_{-0.3}$ &   $2.7^{+0.1}_{-0.1}$ \\
        &     &     &  $6555^{+1139}_{-1245}$& $1.6\times{10^{4}}^{+1.7\times10^4}_{-1.0\times10^3}$ & $0.48^{+0.08}_{-0.03}$ &    $242^{+10.3}_{-11.7}$ &  $2.5^{+2.2}_{-1.4}$ &   $2.0^{+0.8}_{-0.4}$ \\
170202A &  C3 & 108 &  $100^{+4.1}_{-3.2}$ &      $41.0^{+41.1}_{-15.1}$ & $0.65^{+0.09}_{-0.15}$ &        $18.5^{+6.3}_{-4.5}$ &  $0.4^{+0.4}_{-0.1}$ &   $0.6^{+0.1}_{-0.04}$ \\
170113A &  C4 & 273 &  $93^{+0.4}_{-0.5}$  &       $21.1^{+14.3}_{-8.3}$ & $0.44^{+0.06}_{-0.05}$ &           $78.4^{+1.6}_{-1.5}$ &  $0.2^{+0.2}_{-0.1}$ &   $1.8^{+0.3}_{-0.2}$ \\
170113A &  C4 & 273 &  $94^{+0.8}_{-0.9}$  &         $581^{+163}_{-143}$ & $0.34^{+0.01}_{-0.01}$ &             $25.8^{+4.6}_{-5.8}$ &  $6.2^{+1.7}_{-1.5}$ &  $10.3^{+1.4}_{-1.3}$ \\
161219B &  C6 & 737 &  $2480^{+82}_{-79}$  &     $14036^{+1626}_{-2806}$ & $0.42^{+0.01}_{-0.01}$ &         $0.11^{+0.01}_{-0.01}$ &  $5.7^{+0.6}_{-1.1}$ &   $4.8^{+0.7}_{-0.6}$ \\
        &     &     &  $406^{+4.2}_{-4.3}$ &         $315^{+55}_{-45.4}$ & $0.67^{+0.03}_{-0.03}$ &             $0.4^{+0.03}_{-0.03}$ &  $0.8^{+0.1}_{-0.1}$ &  $38.6^{+2.9}_{-2.6}$ \\
161117A & G12 & 560 &  $124^{+0.6}_{-0.6}$ &       $168^{+30.2}_{-23.9}$ & $0.45^{+0.01}_{-0.01}$ &         $364^{+7.9}_{-8.9}$ &  $1.4^{+0.2}_{-0.2}$ &  $42.4^{+1.6}_{-1.5}$ \\
        &     &     &  $7844^{+480}_{-512}$&   $2.7\times{10^{4}}^{+1.3\times10^4}_{-1.2\times10^4}$ & $0.46^{+0.03}_{-0.02}$ &       $27.8^{+4.4}_{-6.3}$ &  $3.5^{+1.6}_{-1.5}$ &   $1.8^{+0.2}_{-0.2}$ \\
161108A &  G6 & 416 &  $148^{+2.8}_{-2.8}$ &           $307^{+82}_{-72}$ & $0.46^{+0.02}_{-0.02}$ &              $37.5^{+2.1}_{-3.3}$ &  $2.1^{+0.5}_{-0.5}$ &   $3.6^{+0.3}_{-0.3}$ \\
        &     &     & $1398^{+1175}_{-557}$&        $420^{+3034}_{-315}$ & $0.37^{+0.13}_{-0.11}$ &           $3.6^{+23.6}_{-3.2}$ &  $0.3^{+1.9}_{-0.2}$ & $32.3^{+294}_{-29.3}$ \\
161017A &  C6 & 1562 &    $409^{+1.2}_{-1.2}$ &        $4630^{+295}_{-557}$ & $0.24^{+0.004}_{-0.003}$ &          $337^{+6.8}_{-6.7}$ & $11.3^{+0.7}_{-1.4}$ &   $1.7^{+0.1}_{-0.1}$ \\
        &     &  &    $201^{+1.4}_{-1.4}$ &        $1712^{+35.1}_{-64}$ & $0.34^{+0.003}_{-0.003}$ &        $88^{+2.8}_{-2.8}$ &  $8.5^{+0.2}_{-0.3}$ &   $3.2^{+0.1}_{-0.1}$ \\
160804A & G12 & 1464 &  $433^{+2.2}_{-2.2}$ &         $4194^{+87}_{-153}$ & $0.31^{+0.004}_{-0.003}$ &            $11^{+0.4}_{-0.4}$ &  $9.7^{+0.2}_{-0.4}$ &   $1.5^{+0.1}_{-0.1}$ \\
        &     &  & $1.5\times{10^{4}}^{+2\times10^3}_{-2\times10^3}$ &   $2.3\times{10^{4}}^{+3.8\times10^4}_{-1.0\times10^4}$ & $0.49^{+0.11}_{-0.06}$ &     $0.7^{+0.4}_{-0.2}$ &  $1.5^{+2.3}_{-0.6}$ &   $0.7^{+0.2}_{-0.1}$ \\
160425A &  G5 & 1004 &     $300^{+0.6}_{-0.6}$ &       $287^{+37.5}_{-30.9}$ & $0.40^{+0.01}_{-0.01}$ &          $34.1^{+0.4}_{-0.4}$ &  $1.0^{+0.1}_{-0.1}$ &   $115^{+6.3}_{-5.4}$ \\
160410AS&  E3 & 77 &  $192^{+32.4}_{-39.3}$ &         $600^{+479}_{-416}$ & $0.45^{+0.07}_{-0.04}$ &           $7.1^{+3.0}_{-2.3}$ &  $3.0^{+2.1}_{-1.9}$ &   $1.1^{+1.4}_{-0.4}$ \\
160228A &  H8 & 95 &     $181^{+2.8}_{-3.3}$ &           $75^{+568}_{-61}$ & $0.39^{+0.13}_{-0.13}$ &            $0.6^{+2.5}_{-2.1}$ &  $0.4^{+1.5}_{-0.3}$ &   $1.0^{+0.5}_{-0.3}$ \\
160227A &  E4 & 859 &    $215^{+0.8}_{-0.8}$ &          $83^{+6.7}_{-6.3}$ & $0.68^{+0.01}_{-0.01}$ &           $186^{+4.9}_{-5.2}$ &  $0.4^{+0.03}_{-0.03}$ &  $16.4^{+0.7}_{-0.7}$ \\
        &     &  &    $422^{+1.1}_{-1.0}$ &       $345^{+16.8}_{-16.0}$ & $0.80^{+0.003}_{-0.004}$ &          $137^{+3.1}_{-3.0}$ &  $0.8^{+0.03}_{-0.03}$ &   $9.9^{+0.7}_{-0.6}$ \\ 
160131A &  E4 &  582 &  $197^{+8.0}_{-11.3}$ &         $871^{+204}_{-268}$ & $0.46^{+0.02}_{-0.01}$ &            $24^{+3.2}_{-4.5}$ &  $4.4^{+0.9}_{-1.2}$ &   $1.9^{+0.6}_{-0.5}$ \\
        &     &  & $6948^{+296}_{-297}$ &   $29744^{+17935}_{-15924}$ & $0.39^{+0.05}_{-0.03}$ &      $6.2^{+1.0}_{-1.1}$ &  $4.3^{+2.5}_{-2.2}$ &   $0.6^{+0.1}_{-0.04}$ \\
151027B &  C3 & 66 &  $3626^{+944}_{-2708}$ &     $3263^{+14307}_{-2535}$ & $0.47^{+0.23}_{-0.13}$ &       $35.7^{+121}_{-13.1}$ &  $1.2^{+3.1}_{-0.5}$ &  $2.3^{+40.7}_{-0.9}$ \\
151027A &  C6 & 909 &    $132^{+0.4}_{-0.4}$ &        $1180^{+121}_{-196}$ & $0.30^{+0.01}_{-0.003}$ &            $49^{+0.82}_{-0.81}$ &  $9.0^{+0.9}_{-1.5}$ &  $35.8^{+1.9}_{-1.9}$ \\
 &   &  &     $307^{+1.9}_{-1.8}$ &        $3038^{+205}_{-380}$ & $0.28^{+0.01}_{-0.004}$ &            $4.9^{+0.2}_{-0.2}$ &  $9.9^{+0.7}_{-1.2}$ &   $1.7^{+0.1}_{-0.1}$ \\
151021A &  C4 & 294 &  $232^{+22.8}_{-11.8}$ &          $185^{+323}_{-77}$ & $0.46^{+0.09}_{-0.10}$ &             $352^{+12}_{-13}$ &  $0.8^{+1.4}_{-0.3}$ &   $1.1^{+0.2}_{-0.1}$ \\
 &   &  &     $129^{+1.0}_{-1.2}$ &        $1046^{+167}_{-242}$ & $0.31^{+0.01}_{-0.01}$ &            $20.5^{+6.5}_{-5.1}$ &  $8.1^{+1.3}_{-1.9}$ &   $4.3^{+0.3}_{-0.4}$ \\
150821A &  G6 & 670 &   $714^{+8.4}_{-7.6}$ &          $633^{+154}_{-87}$ & $0.50^{+0.02}_{-0.03}$ &             $7.0^{+0.73}_{-0.67}$ &  $0.9^{+0.2}_{-0.1}$ &   $1.7^{+0.05}_{-0.04}$ \\
 &   &  &    $1126^{+5.5}_{-5.6}$ &         $264^{+341}_{-148}$ & $0.36^{+0.08}_{-0.07}$ &             $0.53^{+0.1}_{-0.1}$ &  $0.2^{+0.3}_{-0.1}$ &   $0.7^{+0.1}_{-0.1}$ \\
150818A &  E3 &  301  &  $79^{+0.3}_{-0.3}$ &         $421^{+435}_{-291}$ & $0.19^{+0.04}_{-0.02}$ &            $1.03^{+0.1}_{-0.1}$ &  $5.3^{+5.5}_{-3.7}$ &   $3.1^{+0.3}_{-0.3}$ \\
150727A &  E3 &  473 &     $825^{+52}_{-62}$ &       $4008^{+704}_{-1111}$ & $0.45^{+0.01}_{-0.01}$ &           $0.2^{+0.02}_{-0.02}$ &  $4.9^{+0.7}_{-1.2}$ &   $1.2^{+0.1}_{-0.1}$ \\
150314A &  G11& 704 & $1.1\times{10^{4}}^{+4.3\times10^3}_{-1.5\times10^3}$ & $6.4\times{10^{4}}^{+4\times10^6}_{-3.5\times10^4}$ & $0.77^{+0.08}_{-0.06}$ & $98^{+25.1}_{-16.2}$ &  $5.8^{+257}_{-3.0}$ &   $1.5^{+0.2}_{-0.2}$ \\
150206A &  H3 & 813 &    $1911^{+92}_{-167}$ &     $11386^{+5574}_{-5633}$ & $0.32^{+0.03}_{-0.02}$ &          $1722^{+539}_{-199}$ &  $6.1^{+2.7}_{-2.9}$ &   $8.1^{+2.5}_{-0.7}$ \\
 &   &  &  $2362^{+4.5}_{-4.4}$ &     $2.0\times{10^{4}}^{+8.6\times10^3}_{-8.8\times10^3}$ & $0.21^{+0.02}_{-0.01}$ &      $431^{+35.6}_{-31.8}$ &  $8.3^{+3.6}_{-3.7}$ &   $1.6^{+0.2}_{-0.2}$ \\
141221A &  H3 & 69 &  $358^{+33.7}_{-28.4}$ &         $590^{+748}_{-231}$ & $0.50^{+0.06}_{-0.05}$ &           $14.1^{+3.4}_{-2.3}$ &  $1.6^{+1.9}_{-0.5}$ &   $4.4^{+1.1}_{-0.9}$ \\
 &   &  &  $345^{+8.0}_{-7.8}$ &        $431^{+1394}_{-356}$ & $0.33^{+0.16}_{-0.08}$ &             $4.1^{+1.5}_{-1.5}$ &  $1.1^{+3.9}_{-0.9}$ &   $1.6^{+0.6}_{-0.4}$ \\
141121A &  C4 &  882 & $606^{+31.1}_{-25.2}$ &         $244^{+338}_{-146}$ & $0.70^{+0.07}_{-0.13}$ &             $23.7^{+9.2}_{-10.7}$ &  $0.4^{+0.5}_{-0.2}$ &   $2.8^{+0.6}_{-0.4}$ \\
 &   &   &  $6671^{+207}_{-267}$ &   $2.6\times{10^{4}}^{+1.6\times10^4}_{-1.4\times10^4}$ & $0.40^{+0.05}_{-0.02}$ &      $42.2^{+3.6}_{-6.4}$ &  $3.9^{+2.4}_{-2.1}$ &  $13.8^{+3.8}_{-2.5}$ \\
140907A &  C3 & 111 &  $206^{+26.5}_{-18.7}$ &         $601^{+766}_{-437}$ & $0.38^{+0.08}_{-0.05}$ &             $2.6^{+1.1}_{-1.0}$ &  $2.8^{+3.4}_{-2.0}$ &   $3.9^{+2.1}_{-1.2}$ \\
140710A &  G2 & 47 &   $365^{+12.4}_{-15.3}$ &        $140^{+131}_{-35.1}$ & $0.60^{+0.12}_{-0.16}$ &           $0.3^{+0.7}_{-0.5}$ &  $0.4^{+0.3}_{-0.1}$ &  $10.4^{+2.3}_{-1.9}$ \\
140703A &  C5 & 176 &    $120^{+1.6}_{-1.7}$ &       $13.0^{+21.5}_{-3.0}$ & $0.52^{+0.08}_{-0.11}$ &          $58^{+24}_{-14}$ &  $0.1^{+0.2}_{-0.03}$ &   $0.8^{+0.1}_{-0.1}$ \\
140629A &  C5 & 119 &   $202^{+38.7}_{-51}$ &         $229^{+380}_{-102}$ & $0.52^{+0.09}_{-0.07}$ &           $8.5^{+9.0}_{-3.1}$ &  $1.1^{+1.5}_{-0.3}$ &   $1.5^{+0.9}_{-0.4}$ \\
\enddata
\end{deluxetable}

\end{document}